\documentclass{article}
\usepackage{svg}
\usepackage{adjustbox}
\usepackage{bm}
\usepackage[square,numbers,compress]{natbib}
\usepackage{tikz}
\usetikzlibrary{positioning,arrows.meta}
\usetikzlibrary{positioning,arrows,shapes,fit}
\usepackage{PRIMEarxiv}
\usepackage{booktabs} % For professional looking tables (\toprule, \midrule, \bottomrule)
\usepackage{tabularx} % For tables with adjustable width columns (X type)
\usepackage{geometry} % To set page margins if needed (e.g., for wider tables)
\usepackage{amsmath}
\usepackage[utf8]{inputenc} % allow utf-8 input
\usepackage[T1]{fontenc}    % use 8-bit T1 fonts
\usepackage{hyperref}       % hyperlinks
\usepackage{url}            % simple URL typesetting
\usepackage{booktabs}       % professional-quality tables
\usepackage{amsfonts}       % blackboard math symbols
\usepackage{nicefrac}       % compact symbols for 1/2, etc.
\usepackage{microtype}      % microtypography
\usepackage{lipsum}
\usepackage{subfig}
\usepackage{fancyhdr}       % header
\usepackage{graphicx}       % graphics
\graphicspath{{media/}}     % organize your images and other figures under media/ folder
\usetikzlibrary{arrows,positioning,shapes,decorations.pathreplacing}
\newcommand{\state}{\bm{x}}
\newcommand{\action}{\bm{u}}
\newcommand{\traj}{\bm{\tau}}

\newcommand{\cost}{\phi}
\newcommand{\discount}{\gamma}
\newcommand{\statespace}{\mathcal{X}}
\newcommand{\actionspace}{\mathcal{U}}

\title{A Survey and Tutorial of Reinforcement Learning Methods in Process Systems Engineering}
\author{
  Maximilian Bloor \\
  Department of Chemical Engineering\\
  Imperial College London \\
  London\\
  \texttt{\{max.bloor22\}@imperial.ac.uk} \\
   \And
     Max Mowbray\\
  Department of Chemical Engineering\\
  Imperial College London\\
  London\\
  \texttt{\{m.mowbray\}@imperial.ac.uk} \\
  \And
  Ehecatl Antonio del Rio Chanona$^*$\\
  Department of Chemical Engineering \\
  Imperial College London\\
  London\\
  \texttt{\{a.del-rio-chanona\}@imperial.ac.uk} \\
  \And
  Calvin Tsay\thanks{Corresponding Authors}\\
  Department of Computing \\
  Imperial College London\\
  London\\
  \texttt{\{c.tsay\}@imperial.ac.uk}
   }

\begin{document}
\maketitle

\begin{abstract}
Sequential decision making under uncertainty is central to many Process Systems Engineering (PSE) challenges, where traditional methods often face limitations related to controlling and optimizing complex and stochastic systems. Reinforcement Learning (RL) offers a data-driven approach to derive control policies for such challenges. This paper presents a survey and tutorial on RL methods, tailored for the PSE community. We deliver a tutorial on RL, covering fundamental concepts and key algorithmic families including value-based, policy-based and actor-critic methods. Subsequently, we survey existing applications of these RL techniques across various PSE domains, such as in fed-batch and continuous process control, process optimization, and supply chains. We conclude with PSE focused discussion of specialized techniques and emerging directions. By synthesizing the current state of RL algorithm development and implications for PSE this work identifies successes, challenges, trends, and outlines avenues for future research at the interface of these fields.
\end{abstract}

% keywords can be removed
%\keywords{Reinforcement Learning \and Process Systems Engineering}

\section{Introduction}
% Sequential decision problems ever-present across PSE
% RL as a method to solve these problems 
As a field, Process Systems Engineering (PSE) is concerned with managing complex processes that require sequential decision-making under uncertainty~\cite{pistikopoulos2021process}. Consider operating a fed-batch bioreactor: a sequence of decisions regarding the evolution of feed rate over a discrete time horizon influences not just the immediate yield, but the entire trajectory of cell growth and product formation, ultimately impacting final product quality and process economics. From reactor control to supply chain optimization and real-time operations management, PSE practitioners are posed with sequential decision-making problems that demand balancing immediate operational objectives against long-term system performance despite the presence of uncertainty and operational constraints.

The inherent complexity of sequential decision problems presents fundamental challenges to any control methodology. While powerful frameworks such as Model Predictive Control (MPC) leverage an explicit process model for optimization, their effectiveness can be limited by the computational demands of optimizing over discrete control spaces, the formulation of uncertainty, and the tractability of highly nonlinear dynamical models \cite{schwenzer2021review}. 
Reinforcement Learning (RL) is a branch of machine learning focused on how control inputs are selected given state feedback to minimize cumulative costs~\cite{sutton1998reinforcement}. Specifically, RL provides a set of learning rules to abstract an approximately optimal control policy for a given decision process simply through sampling. This has proved powerful for solving a number of challenging problems, with popular recent successes in robotics~\cite{tang2025deep}, autonomous driving~\cite{feng2023dense}, and LLM training~\cite{guo2025survey}; with exploration within PSE over the last decade. By learning directly from interactions with the system or process model, RL algorithms can develop policies without requiring complete closed form descriptions of the underlying processes. When combined with function approximation, RL is emerging as a promising solver for problems across PSE.

Given the growing interest and potential of RL in PSE this paper serves two main goals. First, it provides a tutorial covering the fundamentals and key algorithms. While several valuable reviews have explored RL applications in PSE~\cite{yoo2021reinforcement, faria2022reinforcement,nian2020review,shin2019reinforcement,rajasekhar2025exploring,szatmari2025support} they mostly focus on surveying applications or specific sub-areas rather than providing an in-depth, pedagogical treatment of methods. This paper aims to complement these by exploring a range of techniques including model-free and model-based RL, as well as solvers for specific problem formulations that consider risk aversion, constraints, multiple goals and limited access to further data acquisition. These latter topics, in particular, are common challenges in the PSE domain and have often received comparatively brief discussion or lacked a tutorial focus in prior reviews. Second, this paper surveys existing RL applications across PSE for each of the topics explored. By reviewing the literature, we identify how RL has been used in areas such as process control, scheduling, and supply chains, highlighting successes, common challenges, and current trends.

Combining these tutorial and survey aspects, our contributions are: {\sf{(1)}} an accessible guide to RL methods relevant for PSE, {\sf{(2)}} a summary of current RL applications in PSE, and {\sf{(3)}} an identification of research gaps and promising future directions.

The paper is structured as follows: Section~\ref{sec:problem_setting} presents the problem setting using a general decision process formulation common to PSE, followed by a review of some PSE applications in Section~\ref{sec:applications}. Section~\ref{sec:MDP} introduces the Markov Decision Process (MDP) framework, and Section~\ref{sec:exact_mdp} covers exact MDP solution methods. Sections~\ref{sec:model_free_methods} to \ref{sec:gcrl} form the main tutorial body, through exploration of different RL algorithm classes and review of their applications in PSE. Section~\ref{sec:conc_future} concludes by summarizing findings, discussing challenges in PSE, and identifying areas for future research.
%%%%%%%%%%%%%%%%%%%%%%%%%%%%%%%%%

\section{Problem Setting}\label{sec:problem_setting}
 To build an intuitive understanding of the problem setting, we first present an illustrative example. Following this, we will define the characteristics of the stochastic dynamic systems that form the typical control and optimization tasks in PSE, thereby motivating the need for frameworks such as RL.
\subsection{Illustrative Example}
\label{sec:grid_world_examples}

\begin{figure}[htbp]
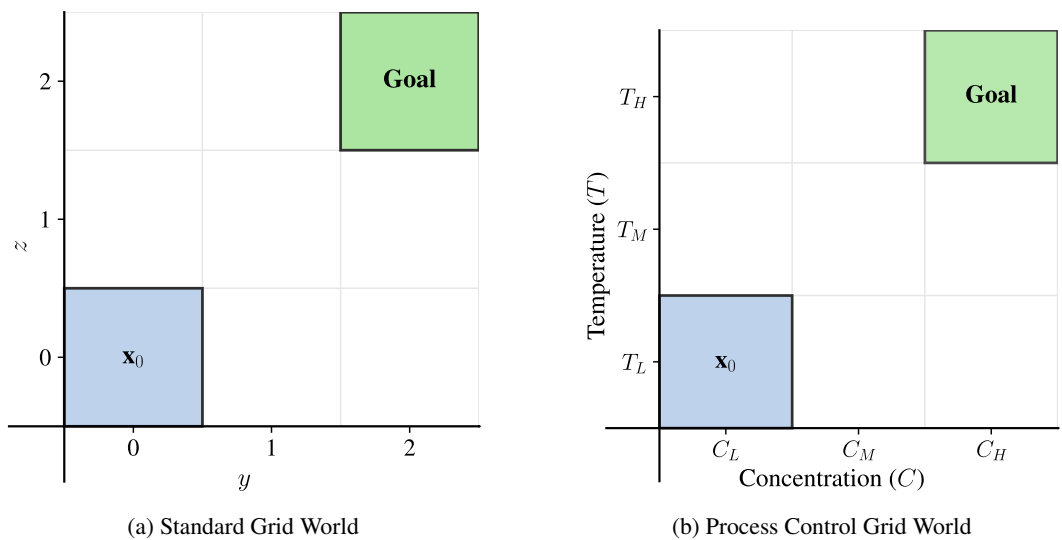

    \centering
    \subfloat[Standard Grid World\label{fig:grid_a}]{%
        \includesvg[width=0.4\textwidth]{figs/grid_world_a.svg}%
    }
    \hspace{1cm}
    \subfloat[Process Control Grid World\label{fig:grid_b}]{%
        \includesvg[width=0.4\textwidth]{figs/grid_world_b.svg}%
    }
    \caption{Illustrative grid world environments.}
    \label{fig:grid_worlds}
\end{figure}

Figure~\ref{fig:grid_worlds} presents two different grid world examples. These examples are presented to help illustrate the components involved when a controller interacts with a system sequentially over time to achieve an objective.

Figure~\ref{fig:grid_a} depicts a prototypical navigation task, e.g., for robotics or autonomous vehicles. The state is given by a $(y,z)$ position, and controls are movements (e.g., up, down). 
The controller is assumed to interact with the environment via a discrete sequence of controls over a discrete time horizon. A control in a state at a given time within the process leads to an uncertain state transition at the next time index.
Specifically, each time a control is input, the state transitions and an associated cost is observed (e.g., energy or time). The objective is to find a policy to reach goal $(y,z) = (2,2)$ from the initial state $(0,0)$ with minimal cumulative cost, potentially whilst avoiding constraint violations. 
Note here that the goal of RL is to learn the optimal \textit{policy}, i.e., the best mapping from the state to the controls, rather than just the best sequence of controls (open loop). 

One can apply similar concepts to process control, as seen in Figure~\ref{fig:grid_b}. In this example, the state comprises the process variables Temperature ($T$) and Concentration ($C$). Controls correspond to setting the control manipulated variables (e.g., adjusting coolant flow or reactant feed), which transitions the process to new states. Transitions can be stochastic, for instance due to process disturbances. A cost function reflects operational goals, penalizing deviations from a target state or energy use. The objective is to find a control policy to reach and maintain this target state whilst minimizing cumulative costs.  While simplified, these grid worlds illustrate key elements of sequential decision-making problems: a system with states evolving through discrete time (or modeled in discrete time) subjected to controls and accumulation of cost. The following sections will mathematically formalize these concepts.

% Propose the master problem then motivate the use of RL compared to where MPC may have been used in the past
\subsection{Control of Stochastic Dynamical Systems}

Previously, we informally described the control of stochastic discrete-time dynamical systems, a framework adopted due to the inherent discrete-time nature of modern instrumentation and control interfaces. 
To establish our notation, we now assume that the underlying dynamical system is characterized by a probabilistic state transition model,
\begin{equation}
   X_{t+1} \sim p(\bm{x}_{t+1} | \bm{x}_t, \bm{u}_t)\,,
\end{equation}
where $\bm{x}_t \in \mathcal{X} \subseteq \mathbb{R}^{n_x}$ represents the system state (e.g. concentration and/or temperature) at time step $t$, and $\bm{u}_t \in \mathcal{U} \subseteq \mathbb{R}^{n_u}$ denotes the vector of control inputs (e.g. coolant flowrate and/or valve position). In the process grid world example (Figure~\ref{fig:grid_b}), an example state could be ($C_L$, $T_H$), and the state space would be the set of all possible ($T, C$)  grid cells. In the same example, a control input $\bm{u}_t$ could be `high coolant flow', or `high reactant feed'. The control space $\mathcal{U}$ is the set of control choices. The notation distinguishes between $X_{t+1}$, the random variable describing the subsequent state, and $\bm{x}_{t+1}$, the specific realization that takes place, assumed to be sampled from the conditional probability density function $p(\cdot|\bm{x}_t,\bm{u}_t)$. 

\begin{figure}[h]
    \centering
    \includesvg[width=0.4\linewidth]{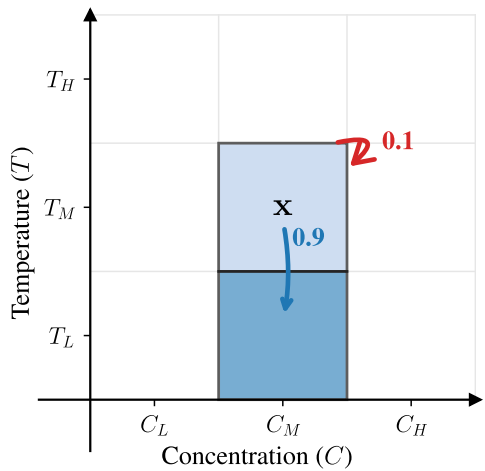}
    \caption{Probabilistic transition in the process control gridworld}
    \label{fig:transition}
\end{figure}

In general, engineers often work with implicit models of probabilistic dynamics, represented by nonlinear state-space models,
\begin{equation}\label{eq:processmodel}
   \bm{x}_{t+1} = f(\bm{x}_t, \bm{u}_t, \bm{d}_t)\,,
\end{equation}
where stochasticity arises from uncertain parameters $\bm{d}_t \in \Omega \subseteq \mathbb{R}^{n_d}$, which may represent process disturbances or model parameters, with $\Omega = \text{supp}(p(\bm{d}))$ defining the support of the corresponding distribution. Crucially, this describes uncertainty within a known model structure, which is distinct from structural uncertainty, which would question whether the model $f(\cdot)$ is an accurate representation of reality in the first place. The structure of these models \eqref{eq:processmodel} is often derived from first-principles reasoning. In the process control grid world example (Figure~\ref{fig:grid_b}), the transition probability defines the probability of moving to a new state given the current cell and control taken. For instance, if the system is at ($T_M, C_M$) and the agent selects `high coolant' as a control input, the environment's state might transition to ($T_L, C_M$) with 90\% probability, but due to a disturbance variable for example, might stay at ($T_M, C_M$) with 10\% probability (illustrated in Figure~\ref{fig:transition}).

For the operation of chemical processes, we seek to learn decision strategies that optimize the distribution of system trajectories given an initial state distribution, $X_0 \sim p_0(\bm{x})$. In other words, we seek to identify a state feedback policy, $\pi = (\pi_0, \ldots, \pi_{T-1})$, where each decision rule maps the state to a distribution over controls $\bm{u}_t \sim \pi_t(\cdot|\bm{x}_t)$. 
The optimal policy $\pi$ minimizes a sum of stage costs (e.g., setpoint error or energy use), defined by a cost function $\phi_t: \mathcal{X} \times \mathcal{U} \times \mathcal{X} \rightarrow \mathbb{R}$, over a discrete time horizon, $\mathcal{T} = \{0,\ldots, T\}$. This finite-horizon stochastic control problem can be formulated as,
\begin{subequations}\label{eq:finite_problem}
\begin{align}
\min_{\pi} \hspace{2mm}&\mathbb{E}_{\bm{\tau}\sim p_\pi} \left[\sum_{t=0}^{T-1} \phi_t(\bm{x}_t,\bm{u}_t, \bm{x}_{t+1})\right] \\
\text{s.t.}\hspace{0.5cm}
& p_\pi(\bm{\tau}) = p_0(\bm{x}_0) \prod_{t=0}^{T-1} \pi_t(\bm{u}_t|\bm{x}_t) p(\bm{x}_{t+1}|\bm{x}_t,\bm{u}_t)\label{eq:state_density}\\
& \bm{u}_t \in \mathbb{U}(\bm{x}_t), \quad \forall t\in \mathcal{T} \setminus \{T\}\\
& \mathbb{P}\left[\bm{x}_t \in \mathbb{X}_t\right] \geq 1-\delta_t, \quad \forall t \in \mathcal{T}\setminus\{0\}
%&t \in \mathcal{T}
\end{align}
\end{subequations}
where $p_\pi(\traj)$ denotes the probability of a trajectory $\traj = (\bm{x}_0, \bm{u}_0, \ldots, \bm{x}_T)$ occurring when policy $\pi$ is followed, as defined in \eqref{eq:state_density}. It should be noted that the Problem~\ref{eq:finite_problem} generalises to familiar control settings if the policy is deterministic ($\pi$ is a dirac delta function) and constraints are fully enforced ($\delta_t =0$). The solution to this stochastic control problem is the determination of an optimal policy, $\pi^*$, which specifies a sequence of decision rules designed to minimize cumulative costs while adhering to several critical considerations. These include accounting for the system's probabilistic dynamics, $p(\state_{t+1}|\state_t,\action_t)$, which, in the context of our process grid world (Figure~\ref{fig:grid_b}), describe the stochastic nature of transitions between states resulting from taking a control in a given state. Furthermore, the policy must respect state-dependent control constraint sets, $\action_t \in \mathbb{U}(\state_t)$, analogous to actuator limitations which define restrictions on control inputs in specific states of the grid world. Finally, joint chance constraints on the states, $\mathbb{P}\left[\state_t \in \mathbb{X}_t\right] \geq 1-\delta_t$, must be satisfied at each time step, reflecting operational, safety, regulatory, or quality requirements, such as maintaining the temperature below a critical value with a high probability, or avoiding obstacle cells in the navigation grid world (Figure~\ref{fig:grid_a}).

We note that, when a detailed mathematical model \eqref{eq:processmodel} is available, solution approaches for problem \eqref{eq:finite_problem} include MPC~\cite{rawlings2020model} and its stochastic~\cite{heirung2018stochastic} and constrained~\cite{mayne2000constrained} variants. 

Within the RL paradigm, the stochastic control problem formulation~\eqref{eq:finite_problem} has motivated the development of specialized subfields, each addressing distinct challenges that arise when learning (optimal) policies from experience. Model-Based RL is motivated to understand system dynamics when the problem requires sample-efficient long-term planning. Constrained RL develops principled approaches for incorporating and satisfying control constraints $\mathbb{U}(\state_t)$ and probabilistic state constraints $\mathbb{P}\left[\state_t \in \mathbb{X}_t\right] \geq 1-\delta_t$\footnote{It is worth noting that the CMDP framework typically focuses on satisfying constraints in expectation, which is well suited to stochastic approximation schemes, but typically insufficient for most engineering systems.}.

Safe RL constitutes a broader and distinct subfield that addresses the critical need for safety during the learning process itself by enforcing pathwise constraints and preventing unsafe exploration. Goal-Conditioned RL (Section~\ref{sec:gcrl}) tackles the practical requirement for adaptable policies that can handle varying control objectives, enabling a single learned policy to generalize across multiple operational targets or setpoints. Additional RL branches further specialize in addressing other fundamental aspects of stochastic control, such as partial observability and the distributional nature of costs, allowing for risk-sensitive decision-making.

In addition to these problem-specific challenges, large-scale stochastic control problems appear frequently in PSE. The dimensionality of state and control spaces presents significant tractability issues. A distillation column may have hundreds of state variables, while a chemical plant with multiple interconnected units involves even larger state spaces. This ``curse of dimensionality'' motivates hierarchical and distributed control architectures. Furthermore, classical control objectives such as offset-free setpoint tracking and closed-loop stability guarantees remain critical in PSE practice but are not inherently addressed by standard RL formulations. These considerations necessitate hybrid approaches that integrate RL with established control-theoretic concepts to achieve both scalability and the performance guarantees required for industrial deployment.

%%%%%%%%%%%%%%%%%%%%%%%
\section{Applications}
\label{sec:applications}
\begin{figure}[h!]
    \centering
    \includesvg[width=1\linewidth]{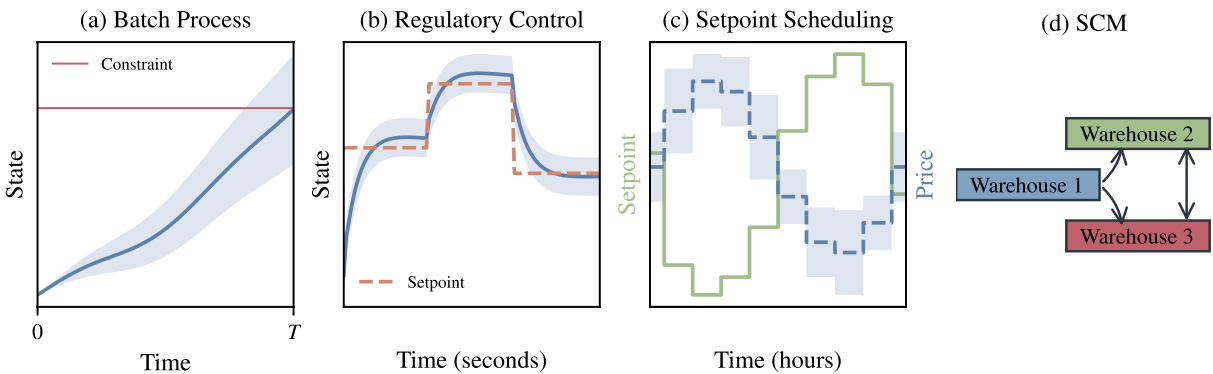}
    \caption{Illustration of applications of RL in Process Systems Engineering}
    \label{fig:RL_App}
\end{figure}

Before describing RL methodologies, we first briefly review some applications in PSE. 
RL has been applied across PSE domains, offering solutions for problems at different scales and layers of the control and optimization hierarchy. This section reviews the application of RL across four key domains: batch process control, regulatory control, supply chain management (SCM), and setpoint scheduling (Figure~\ref{fig:RL_App}), highlighting how different RL methodologies are matched to the specific challenges of each area. 
For a more comprehensive review of applications, the interested reader is referred to recent reviews~\cite{yoo2021reinforcement, faria2022reinforcement,nian2020review,shin2019reinforcement,rajasekhar2025exploring,szatmari2025support}. 
Implementations of prototypical applications are available in PSE-focused adaptations of the popular gym environment, such as PC-gym~\cite{bloor2024pc}, OR-gym~\cite{hubbs2020or}, and SafeOR-gym~\cite{ramanujam2025safeor}. 

\subsection{Batch Process Control}
Batch processes are integral to the manufacturing of high-value products, such as pharmaceuticals and specialty chemicals. These operations are characterized by their finite-horizon nature, where the principal objective is the optimization of an end-of-batch metric, for instance, final product quality and yield~\cite{rippin1993batch}. Applications are widespread, ranging from crystallization~\cite{benyahia2021control} to polymerization systems~\cite{ma2019continuous}. The dominant technical challenges in this domain are the management of significant process uncertainties, and highly non-linear dynamics. Traditionally, model-based techniques are used in batch process control, but these require time-consuming and expensive model construction compared to model-free RL, which builds control policies through interaction alone. 

\citet{ma2019continuous}, for instance, applied the Deep Deterministic Policy Gradient (DDPG) algorithm to a non-linear, semi-batch polymerization system. This required adaptations, including augmenting the state with historical data to overcome process delays and formulating a tailored reward function. For bioprocessing, which involves stochasticity and plant-model mismatch, \citet{petsagkourakis2020reinforcement} integrated policy gradient methods with recurrent neural networks for batch-to-batch optimization. They introduced a two-stage methodology: an approximate model is first used for extensive offline policy training. This policy then undergoes online training via transfer learning, which allows the policy to adapt to the true plant dynamics. The approach outperformed a Nonlinear MPC by better managing system uncertainty and model inaccuracies.

In addition to directly managing the control of batch process dynamics, ensuring process robustness against uncertainty is a critical requirement. \citet{byun2020robust} utilized Proximal Policy Optimization (PPO) to develop dual-control agents. These agents concurrently optimize the process while actively identifying system parameters, thereby yielding robust policies capable of adapting to operational variability.

A significant impediment to the application of RL in batch processing is the prohibitive cost of data acquisition. One strategy to mitigate this issue is to leverage existing process knowledge through inverse RL (IRL) \citep{ng2000algorithms, mowbray2021using}. \citet{anandan2022optimal} demonstrated this approach by inferring the operational policy of a crystallization process from an expert Nonlinear Model Predictive Control (NMPC) scheme. Addressing the problem of sample efficiency from a direct algorithmic standpoint, \citet{bloor2025gaussian} proposed Gaussian Process Q-Learning, a method specifically designed for finite-horizon problems where data collection is resource-intensive. This algorithm replaced the state-action value approximator\footnote{This refers to some function approximation to the expected cost associated with a state-control pair under a given policy.}, which is typically a neural network, with a Gaussian process. The Gaussian process is built using a subset of the available dataset, selected using a determinantal point process \cite{kulesza2012determinantal}, which builds a subset which balance diversity and quality. The algorithm shows improved sample efficiency compared to an algorithm leveraging deep neural networks, known as the Deep Q Network (DQN) \cite{mnih2013playingatarideepreinforcement}. 

\subsection{Regulatory Control}
At the foundational layer of the automation hierarchy, regulatory control is responsible for maintaining critical process variables at their setpoints. While Proportional-Integral-Derivative (PID) controllers dominate this area, their performance depends on effective tuning, which is a challenge for complex or time-varying systems. RL offers a model-free approach to automatically learn and adapt feedback control policies for such systems, potentially reducing or eliminating the need for manual tuning while maintaining performance under changing conditions.

Rather than replacing the PID controller, many RL approaches seek to augment it. Some strategies employ meta-RL, where an agent is trained offline on a wide variety of simulated processes to enable the rapid online tuning of controllers for new, unseen systems, a technique effectively implemented by \citet{mcclement2022meta}. Another ``control-informed'' approach, proposed by \citet{bloor2025cirl}, embeds the PID structure directly into the RL agent’s policy network, leveraging control-theoretic knowledge to improve performance, robustness, and data efficiency.

\subsection{Setpoint Scheduling}
Sitting above the regulatory layer, setpoint scheduling adjusts process setpoints to meet economic objectives. These algorithms make decisions with respect to uncertain cost and demand forecasts which can be difficult to model or if they are modeled are timely to solve. Instead, RL can build a control policy from interaction alone which can adapt to changing plant and market conditions during operation.
% AC comment on contextual bandit
It is important to note that the appropriate framework depends on the problem's temporal characteristics. Steady-state real-time optimization problems, where process dynamics are much faster than economic conditions change, reduce to sequences of independent optimization decisions that can be solved using Bayesian optimization~\cite{del2021real}. However, when process dynamics and economic factors (such as electricity prices or demand fluctuations) operate on similar timescales, the full sequential decision-making framework of RL becomes necessary to capture temporal dependencies and long-term consequences of current actions. Intermediate cases with limited temporal coupling can be addressed through non-myopic Bayesian optimization techniques~\cite{cheonearl,astudillo2021bayesian}. The reader is referred to \citet{tsay2019110th} for an overview of these scale-bridging formulations.

RL offers a compelling model-free and adaptive alternative to traditional model-based setpoint scheduling. This concept extends to dynamic optimization, where RL has been applied to complex problems such as reactor start-up by \citet{machalek2020dynamic} and finding optimal temperature profiles by \citet{kannan2023efficient}.

A significant trend is the development of hybrid approaches that merge RL with traditional methods. To combat the persistent issue of model-plant mismatch, \citet{zhang2023enhanced} integrated RL with Economic MPC, using the RL agent to continuously update model parameters to reflect real-world changes such as catalyst deactivation, thereby enhancing the long-term performance of the model-based controller. \citet{bloor2025hierarchicalrlmpcdemandresponse} designed a hierarchical framework where a high-level RL agent sets economically optimal schedules for a lower-level MPC system that ensures robust constraint satisfaction.
% Add a sentence on clarifying RTO as contextual bandits 
% Steady-state RTO is contextual bandit, if dynamics are considered then it is RL. Need solid reason for including dynamics i.e. electricity price and dynamics are on similar time scales. 

\subsection{Supply Chain Management}
Supply chain management involves a web of sequential decisions under significant uncertainty, making it well-suited for RL. In the single-agent paradigm, a centralized controller makes decisions for the entire system. Here, critical challenges include ensuring constraint satisfaction and handling incomplete information. To address this, \citet{BURTEA2024108518} developed a constrained Q-learning algorithm that guarantees controls adhere to known operational limits. To handle partial observability, \citet{rangel2024recurrent} used a Deep Recurrent Q-Network (DRQN) to create an online scheduler for batch plants that makes effective decisions from limited historical data.

However, since real-world supply chains are decentralized, Multi-Agent RL (MARL) offers a more natural framework. For instance, \citet{li2023multi} and \citet{mousa2024analysis} both applied MARL based on PPO variants to develop effective, decentralized ordering policies for inventory management, demonstrating superior performance over traditional heuristics. To better capture the system's structure, some frameworks incorporate Graph Neural Networks (GNNs), as shown by \citet{kotecha2025leveraging}, allowing agents to learn policies that are explicitly aware of the supply chain’s topology. While this work will not explicitly discuss MARL algorithms, we refer the reader to \cite{albrecht2024multi} for further reading.

%%%%%%%%%%%%%%%%%%%%%%%%%%%%%%%%%
\section{The Markov Decision Process Framework}\label{sec:MDP}
% Propose distributional problem
The stochastic control problem described in \eqref{eq:finite_problem}, which inherently relies on the Markov assumption, can be formalized within the MDP framework, which provides the mathematical foundation for the development and analysis of many RL algorithms. An MDP is characterized by a tuple $\mathcal{M} = \langle \statespace, \actionspace, p, \cost, \discount\rangle$, encompassing the state space, control space, transition dynamics, cost function, and discount factor, respectively. The discount factor, $\discount \in [0,1]$, plays a role in weighting future costs against immediate ones. 
For finite-horizon problems as formulated herein, $\discount$ is frequently set to $1$, implying that all costs within the horizon are valued equally without diminishing future costs, whereas for infinite-horizon scenarios ($T=\infty$), $\discount < 1$ is typically employed to ensure convergence of the cumulative cost term and to represent a preference for earlier costs. 

Central to the MDP framework is the the Markov assumption, which posits that the future state of the system depends only on the current state and control, and is therefore conditionally independent of all previous states and controls. The significance of this property is that it simplifies the decision-making process substantially; without it, an agent would potentially need to consider the entire history of observations and controls to determine an optimal decision, leading to intractable models of history-dependent policies. It should be mentioned that Ordinary Differential Equations without time delays, such as those commonly found in PSE problems, also follow this property.% AC comment on ODEs/Markov. 
This Markov assumption treats the current state $\state_t$ as sufficient to encapsulate all relevant information necessary for future predictions and controls. This property underpins the formulation of the unconstrained problem statement,
\begin{align}\label{eq:mdp}
\min_{\pi} \hspace{2mm}&\mathbb{E}_{\traj\sim p_\pi} \left[\sum_{t=0}^{T-1} \discount^t\cost_t(\state_t,\action_t, \state_{t+1})\right] \nonumber \\
\text{s.t.}\hspace{0.5cm}
& p_\pi(\traj) = p_0(\state_0) \prod_{t=0}^{T-1} \pi_t(\action_t|\state_t) p(\state_{t+1}|\state_t,\action_t)\\
& \action_t \in \mathbb{U}(\state_t) \quad\forall t \in \mathcal{T}\setminus{T}\nonumber\\
&t \in \mathcal{T}.\nonumber
\end{align}

\subsection{Infinite Horizon Problems}
The infinite horizon problem extends the stochastic control framework (Problem \ref{eq:finite_problem}) to scenarios requiring consideration of system behavior over indefinite time periods, such as those commnonly found in continuous manufacturing. The framework is also commonly applied in other RL domains where there is no obvious finite horizon length, making it easier to apply an infinite horizon and use $\discount$ to specify performance preferences. This formulation can be expressed as,
\begin{align}\label{prob:inf_horizon}
\min_{\pi} \hspace{2mm}&\mathbb{E}_{\bm{\tau}\sim p_\pi} \left[\sum_{t=0}^{\infty} \gamma^t \phi(\bm{x}_t,\bm{u}_t, \bm{x}_{t+1})\right]  \nonumber\\
\text{s.t.}\hspace{0.5cm}
& p_\pi(\bm{\tau}) = p_0(\bm{x}_0) \prod_{t=0}^{\infty} \pi(\bm{u}_t|\bm{x}_t) p(\bm{x}_{t+1}|\bm{x}_t,\bm{u}_t)\\
& \bm{u}_t \in \mathbb{U}(\bm{x}_t)\nonumber\\
&t \in \{0\ldots\infty\}\nonumber,
\end{align}
with discount factor $\gamma \in [0,1)$. A fundamental concept in infinite horizon problems, given that the transition function and cost function are stationary, is that the solution policy $\pi^*$ is stationary, which means that it does not vary in time. Under a stationary policy and dynamics, the state distribution may converge toward a steady-state distribution as $t \rightarrow \infty$, provided the induced Markov chain satisfies certain regularity conditions. For further reading we direct the reader to~\cite{puterman2014markov}.  

\subsection{Partially Observable Markov Decision Processes}
% Check definition of observability in both communities
Process systems engineering applications frequently include situations where complete state information is unavailable to controllers. 
% AC comment on the comparison of obs in control & RL
For instance, state estimation techniques are a key component of many control frameworks~\cite[Chapter 4]{rawlings2020model}.
While classical control theory addresses this through observer design and state estimation, the RL community approaches similar challenges through the Partially Observable Markov Decision Process (POMDP) framework that provides a framework for decision making under this lack of information.  In the scenario introduced in Section \ref{sec:grid_world_examples}, if the measurements of temperature and concentration were instead from a noisy sensor or not measured directly, this would correspond to a so-called `partially observable' environment. Specifically, a POMDP extends the MDP formalism with an observation model, resulting in an extended tuple $\mathcal{P} = \langle \mathcal{X}, \mathcal{U}, \mathcal{O}, p, q, \phi, \gamma \rangle$. The observation space $\bm{o}_t \in \mathcal{O} \subseteq \mathbb{R}^{n_o}$ captures the set of possible state observations, which are assumed to be generated by the probability density, $q(\bm{o}_t|\bm{x}_t)$.

The history, denoted as $\bm{H}_t = (\bm{o}_0, \bm{u}_0, \bm{o}_1, \bm{u}_1, \ldots, \bm{u}_{t-1}, \bm{o}_t)$, represents the complete sequence of observations and controls available to the actor up to time $t$. This history captures all information available to the actor in terms of previous measurements and all control actions taken,. The policy $\pi(\bm{u}_t|\bm{H}_t)$ maps this observation history to control actions, this presents a computational challenge given the growing dimensionality of the history $\bm{H}_t$. The POMDP problem is therefore defined as,

% construct POMDP problem
\begin{align}
\min_{\pi} \hspace{2mm}&\mathbb{E}_{\bm{\tau}_{po}\sim p_\pi} \left[\sum_{t=0}^{T-1} \phi_t(\bm{x}_t,\bm{u}_t, \bm{x}_{t+1})\right]  \nonumber\\
\text{s.t.}\hspace{0.5cm}
& p_\pi(\bm{\tau}_{po}) = p_0(\bm{x}_0)q(\bm{o}_0\mid\bm{x}_0) \prod_{t=0}^{T-1} \pi(\bm{u}_t|\bm{H}_t) p(\bm{x}_{t+1}|\bm{x}_t,\bm{u}_t)q(\bm{o}_{t+1}\mid 
\bm{x}_{t+1})\label{eq:POMDP}\\
& \bm{u}_t \in \mathbb{U}(\bm{x}_t)\nonumber \quad \forall t \in \mathcal{T} \setminus \{T\}\\
& t \in \mathcal{}{T}\nonumber
\end{align}
with $\bm{\tau}_{po} = (\bm{x}_0, \bm{o}_0, \bm{u}_0, \ldots, x_{T})$. The fundamental challenge in POMDPs arises from having to make control decisions based on incomplete information. While the system evolves according to the state transition model $p(\bm{x}_{t+1}|\bm{x}_t,\bm{u}_t)$, the controller only has access to observations generated through $q(\bm{o}_t|\bm{x}_t)$. To address this challenge, the concept of belief states is usually introduced. A belief state is a probability distribution over possible true states, summarizing all past observations and actions. This converts the POMDP into a  \textit{belief} MDP where decisions are made based on these probability distributions rather than hidden states, restoring the Markov property needed for standard solution methods~\cite{wiering2012reinforcement}. We direct the interested reader to ~\cite{krishnamurthy2016partially} for a more comprehensive overview of POMDPs. 

\section{Exact MDP Solution Strategies}\label{sec:exact_mdp}

The Markov Decision Process framework introduced in Section~\ref{sec:MDP} provides a formal basis for sequential decision making under uncertainty. Provided that one is given the complete model of the MDP, comprising the state space $\mathcal{X}$, control space $\mathcal{U}$, transition dynamics $p(\bm{x}_{t+1}|\bm{x}_t,\bm{u}_t)$, cost function $\phi(\bm{x}_t,\bm{u}_t, \bm{x}_{t+1})$, and discount factor $\gamma$, classical methods from Dynamic Programming (DP) can, in principle, determine the optimal policy $\pi^*$~\cite{bertsekas2024course, bellman1966dynamic}. This optimal policy minimizes the expected cumulative discounted cost objective, which for the finite horizon case is typically defined as shown in Problem~\ref{eq:mdp}. Given discrete (or discretized) state and controls spaces, dynamic programming approaches achieve this by computing optimal value functions via enumeration of all possible state and control pairs, from which the optimal policy can be readily derived. We briefly discuss value functions and DP algorithms below.

\subsection{Optimal Value Functions and Bellman Equations}\label{sec:valuefunctions}
\begin{figure}
    \centering
    \includegraphics[width =0.9\linewidth]{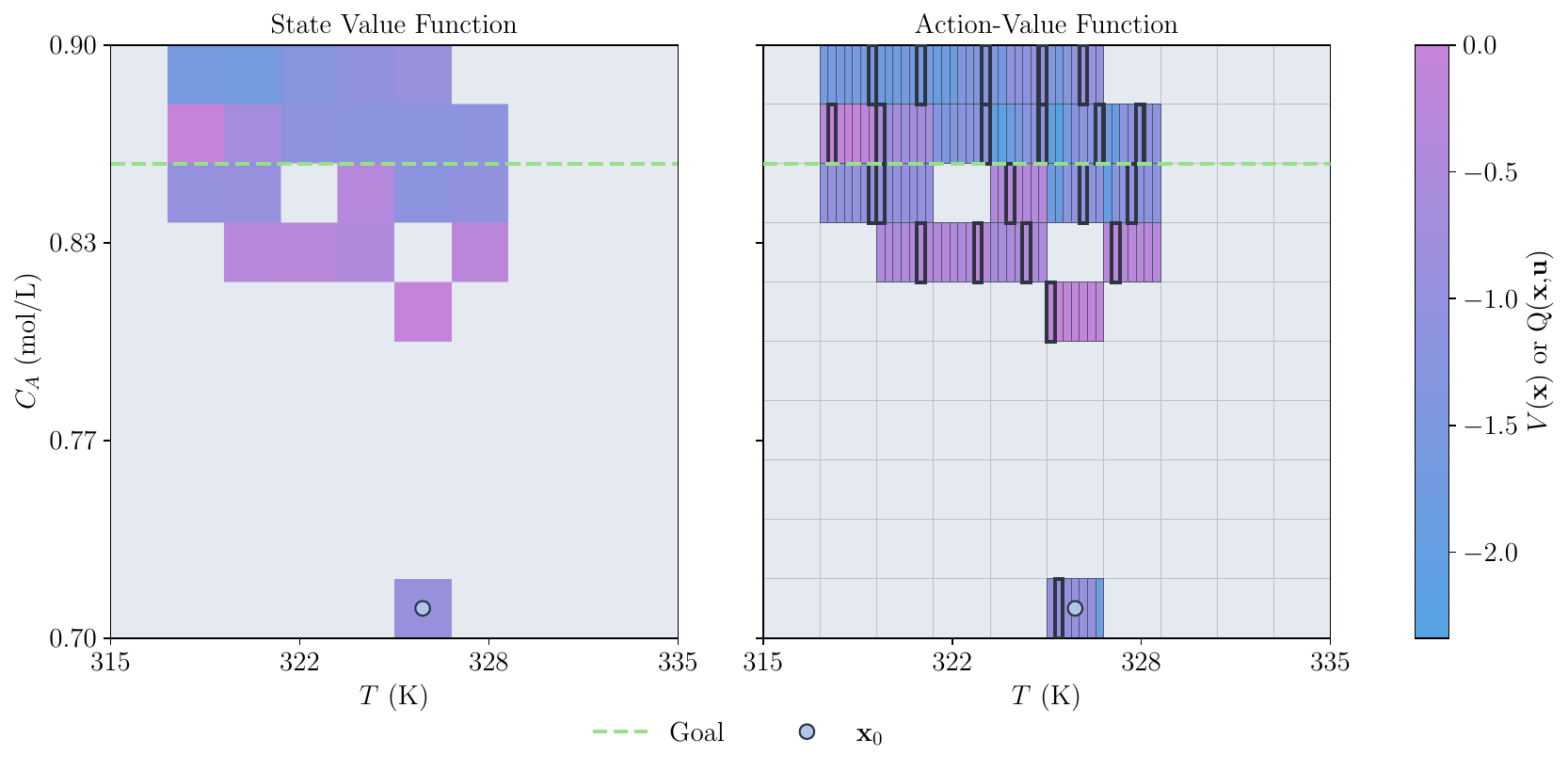}
    \caption{Value function (left) and Q-function (right) for process control gridworlds. The key difference between the two plots is the extra degree of freedom the Q-function has compared to the value function that allows the comparison of different controls. The right hand side plot displays the Q-values of state-action pairs where states are the grid squares and the actions are bars within each grid square.}
    \label{fig:Q_V_function}
\end{figure}

%AC comment on chem eng example
Central to dynamic programming are the concepts of optimal value functions, which quantify the minimum expected future cost achievable from a given state or state-action pair\footnote{The following presentation is restricted to infinite horizon problems. The ideas can be applied within the context of finite horzion problems; however the corresponding value functions become time dependent and exact solution approaches vary.}. For example, one may imagine the value of a state far from the setpoint being lower than the states closer to the setpoints due to the cost of moving the system towards the setpoint. Similarly, the state-action value for a state and an action that brings the system to its setpoint is lower than that of a pair that moves the system away, since the future cost will be greater. The optimal state-value function, denoted $V^*(\bm{x})$, represents the minimum expected cumulative discounted cost, starting from state $\bm{x}$ and thereafter following the optimal policy $\pi^*$, given by
\begin{equation*}
V^*(\bm{x}_t) = \min_{\pi} \mathbb{E}_{p_\pi}\left[\sum_{k=t}^\infty \gamma^{k-t} \phi(\bm{x}_{k}, \bm{u}_{k}, \bm{x}_{k+1}) \right].
\end{equation*}
% AC linking comment
The value function above denotes the value of a state. It may be convenient to allow the comparison of different actions from this state. This motivates the definition of the action-value function. 
The optimal action-value function (see Figure \ref{fig:Q_V_function}), denoted $Q^*(\bm{x}_t, \bm{u}_t)$, represents the minimum expected cumulative discounted cost incurred by taking control $\bm{u}_t$ in state $\bm{x}_t$ and then subsequently following the optimal policy $\pi^*$, calculated as
\begin{equation*}
Q^*(\bm{x}_t, \bm{u}_t) = \mathbb{E}_{\bm{x}_{t+1} \sim p(\cdot|\bm{x}_t,\bm{u}_t)} \left[ \phi(\bm{x}_t, \bm{u}_t, \bm{x}_{t+1}) + \gamma V^*(\bm{x}_{t+1}) \right].
\end{equation*}

Notice that $Q^*$ and $V^*$ only differ in that $Q^*(\bm{x}_t,\bm{u}_t)$ introduces the value of taking action $\bm{u}_t$ rather than exclusively following $\pi^*$~\cite{howard1960dynamic}. The two optimal value functions are therefore related through $V^*(\bm{x}) = \min_{\bm{u}} Q^*(\bm{x}, \bm{u})$. 
These optimal value functions are uniquely characterized by the Bellman optimality equations~\cite{bellman1957markovian}. The optimal state-value function satisfies the Bellman optimality equation
\begin{equation}\label{eq:bellman_optimality}
V^*(\bm{x}_t) = \min_{\bm{u}}  \mathbb{E}_{\state_{t+1} \sim p(\cdot \mid \state_t, \action_t)} \left[ \phi(\bm{x}_t, \bm{u}_t, \bm{x}_{t+1}) + \gamma V^*(\bm{x}_{t+1}) \right] .
\end{equation}
Similarly, the optimal action-value function satisfies
\begin{equation*}
Q^*(\bm{x}_t, \bm{u}_t) = \mathbb{E}_{\state_{t+1}\sim p(\cdot \mid \state_t, \action_t)}  \left[ \phi(\bm{x}_t, \bm{u}_t, \bm{x}_{t+1}) + \gamma \min_{\bm{u}_{t+1}} Q^*(\bm{x}_{t+1}, \bm{u}_{t+1}) \right].
\end{equation*}
These equations express a fundamental property of optimality: the value of a state (or state-action pair) under an optimal policy must equal the expected immediate cost plus the discounted value of the resulting successor state (or optimal action-value from the successor state), minimized over all possible controls. 

Once either $V^*$ or $Q^*$ is known, an optimal deterministic policy $\pi^*(\bm{x})$ can be determined by selecting the control that minimizes the expected future cost according to the Bellman optimality equation at each state according to
\begin{equation*}
\pi^*(\bm{x}_t) \in \arg\min_{\bm{u}_t}  \mathbb{E}_{\state_{t+1}\sim p(\cdot \mid \state_t, \action_t)}\left[ \phi(\bm{x}_t, \bm{u}_t, \bm{x}_{t+1}) + \gamma V^*(\bm{x}_{t+1}) \right],
\end{equation*}
or equivalently, using the optimal action-value function via
\begin{equation}\label{eq:opt_pol_q}
\pi^*(\bm{x}_t) \in \arg\min_{\bm{u}_t} Q^*(\bm{x}_t, \bm{u}_t).
\end{equation}

However, notice that computing either $V^*$ or $Q^*$ (the optimal value functions, as opposed to $V^\pi$ and $Q^\pi$ for a specific policy $\pi$) themselves require the optimal policy $\pi^*$, giving rise to a circular relationship: one can compute $V^*$ or $Q^*$ from $\pi^*$, and vice versa.
\subsection{Dynamic Programming Algorithms}

The above relationship motivates using iterative numerical methods to compute $V^*$ (or $Q^*$) and $\pi^*$ and solve the MDP. 
Dynamic programming precisely provides iterative algorithms to solve the Bellman optimality equations and find the optimal value functions for infinite horizon problems. The primary methods are Value Iteration and Policy Iteration, which can be understood as the repeated application of different control mapping operators. To return to the setpoint tracking example, this allows the iterative construction of the optimal state (or state-action) values if one has knowledge of the system dynamics from which a policy can be derived, or, alternatively, iteratively construct the optimal policy to reach the system's setpoint. The following presentation considers infinite horizon problems; however simple modifications can be made to make the description amenable to solving finite horizon problems.

Policy Iteration (PI)~\cite{howard1960dynamic} alternates between two steps: policy evaluation using the Bellman expectation contraction operator $\mathcal{T}^{\pi}$ and policy improvement using the Bellman optimality contraction operator $\mathcal{T}^*$. First, in policy evaluation, the state-value function $V^{\pi_k}(\bm{x}_t)$ for the current deterministic policy $\pi_k$ is computed by iteratively applying the Bellman expectation operator until convergence:
\begin{equation}\label{eq:bellman_expectation}
V^{\pi_k}_{j+1}(\bm{x}_t) = \mathcal{T}^{\pi_k}V^{\pi_k}_j(\bm{x}_t) = \mathbb{E}_{\state_{t+1}\sim p(\cdot \mid \state_t, \action_t)} \left[ \phi(\bm{x}_t, \pi_k(\bm{x}_t), \bm{x}_{t+1}) + \gamma V^{\pi_k}_j(\bm{x}_{t+1}) \right].
\end{equation}
The Bellman expectation operator $\mathcal{T}^{\pi_k}$ is a contraction mapping, which guarantees convergence of the iterative process to the unique fixed point $V^{\pi_k}$. For finite state spaces, this corresponds to solving a system of $|\mathcal{X}|$ linear equations; for continuous spaces, this represents an integral equation that is typically solved by numerical approximation. Second, in policy improvement, a new policy $\pi_{k+1}$ is generated through a greedy optimization step with respect to the computed value function $V^{\pi_k}$:
\begin{equation*}
\pi_{k+1}(\bm{x}_t) \leftarrow \arg\min_{\bm{u}_t}  \mathbb{E}_{\state_{t+1}\sim p(\cdot \mid \state_t, \action_t)}\left[ \phi(\bm{x}_t, \bm{u}_t, \bm{x}_{t+1}) + \gamma V_j^{\pi_k}(\bm{x}_{t+1}) \right] .
\end{equation*}
These two steps are repeated until the policy stabilizes and converges to $\pi^*$. 

Value Iteration (VI)~\cite{shapley1953stochastic} directly applies the both the Bellman optimality operator $\mathcal{T}^*$  and the Bellman expectation operator $\mathcal{T}^{\pi_k}$ iteratively. Starting with an arbitrary initial value function $V_0(\bm{x})$, VI proceeds by iteratively applying the update rule
\begin{equation*}
V_{k+1}(\bm{x}_t) = \mathcal{T}^*\mathcal{T}^{\pi_k}V_k(\bm{x}_t) = \min_{\bm{u}_t } \mathbb{E}_{\state_{t+1}\sim p(\cdot \mid \state_t, \action_t)} \left[ \phi(\bm{x}_t, \bm{u}_t, \bm{x}_{t+1}) + \gamma V_k(\bm{x}_{t+1}) \right].
\end{equation*}
This iteration is repeated for $k=0, 1, 2, \dots$ until the value function converges. The sequence $\{V_k\}$ is guaranteed to converge to $V^*$ as $k \to \infty$, based on the fixed point theorem~\cite{sutton1998reinforcement}. Notably, Value Iteration can be viewed as a special case of Policy Iteration, where the policy evaluation step (Equation~\ref{eq:bellman_expectation}) is applied only once ($j=0$) before proceeding to policy improvement. For further reading on dynamic programming we refer the reader to~\cite{bertsekas2012dynamic}.

\subsection{Limitations and the Motivation for Reinforcement Learning}

Despite the theoretical guarantees provided by exact DP methods, their practical application in PSE and many other engineering domains is severely hampered by several fundamental obstacles. Firstly, these algorithms generally demand complete knowledge of the system's transition dynamics $p(\bm{x}_{t+1}|\bm{x}_t,\bm{u}_t)$ and cost function $\phi(\bm{x}_t, \bm{u}_t, \bm{x}_{t+1})$, which are often unavailable or intractable for complex industrial processes. Secondly, exact DP methods suffer from the ``curse of dimensionality'': exact DP is intractable for continuous state and control spaces, and, even for large discrete spaces, the computational requirements scale exponentially with dimensionality~\cite{littman2013complexity}, rendering it infeasible for the high-dimensional problems typical of PSE applications. 

These fundamental limitations have motivated the development of approximate solution methods that can operate without perfect model knowledge and/or improve computational tractability of DP. Approximate Dynamic Programming (ADP) methods \citep{bertsekas1996neuro} represent one such approach, approximating the value function or policy using parametric forms (e.g., neural networks or basis functions) to tackle the curse of dimensionality. While ADP techniques have shown success in various process control problems \cite{lee2010approximate, lee2005approximate}, they still typically require some form of model knowledge for planning or simulation.

RL extends these approximation concepts further by eliminating the model requirement entirely, learning optimal policies directly from interactions with an environment, while leveraging the core optimality principles established by dynamic programming theory. The model-free RL methods presented in the following section build upon the Bellman equation framework introduced here (though this is not necessary for value functions in general), but replace the explicit model-based computations with data-driven learning algorithms that can discover optimal policies through experience.
The interested reader is referred to \cite{powell2007approximate} for a comprehensive description of ADP methods, and to \cite{powell2022reinforcement} for extended discussions of the methodological similarities between ADP and RL.

%%%%%%%%%%%%%%%%%%%%%%%%%%%%%%%%%
\section{Model-free Methods}\label{sec:model_free_methods}
Building upon the dynamic programming foundations established in the previous section, model-free RL methods address the practical limitations of exact DP approaches, while preserving (some) theoretical optimality principles. Rather than requiring a complete model of system dynamics, these methods adapt the core Bellman equation framework to operate with sampled experience data, i.e., samples gathered via direct interaction with the environment.

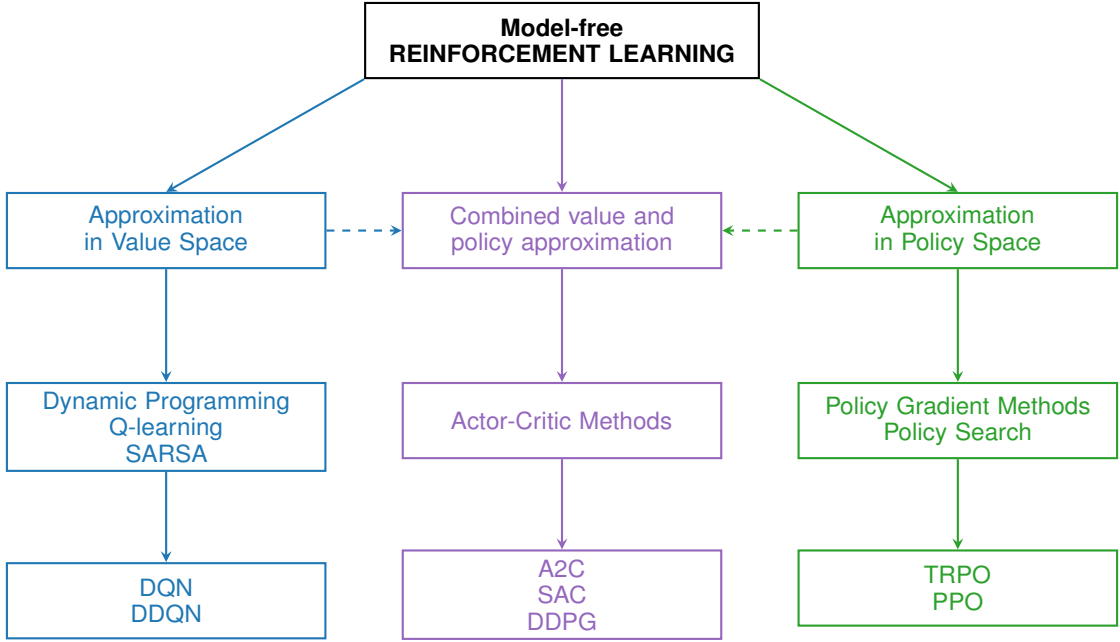
\begin{figure}
\centering
% Define Tableau colors
\definecolor{tableaublue}{HTML}{1f77b4}
\definecolor{tableaugreen}{HTML}{2ca02c}
\definecolor{tableaupurple}{HTML}{9467bd}
\definecolor{tableaugray}{HTML}{7f7f7f}
\begin{tikzpicture}[
    node distance = 2cm,
    box/.style = {
        rectangle, draw, text width=4cm, minimum height=1cm, align=center, font=\sffamily\small
    },
    bluebox/.style = {
        box, draw=tableaublue, thick, text=tableaublue
    },
    redbox/.style = {
        box, draw=tableaugreen, thick, text=tableaugreen
    },
    purplebox/.style = {
        box, draw=tableaupurple, thick, text=tableaupurple
    },
    blackbox/.style = {
        box, draw=black, thick, text=black
    },
    arrow/.style = {
        ->, thick, >=stealth
    }
]
% Main heading
\node[blackbox, text width=5cm, font=\bfseries\sffamily\small] (rl) {Model-free\\ REINFORCEMENT LEARNING\\[0.1cm]};
% First level nodes
\node[bluebox, below left=1.5cm and 0.5cm of rl] (value) {Approximation in Value Space};
\node[redbox, below right=1.5cm and 0.5cm of rl] (policy) {Approximation in Policy Space};
\node[purplebox, below= 1.5cm of rl](actorcriticapprox){Combined value and policy approximation};
\node[purplebox, below=1.5cm of actorcriticapprox] (actorcritic) {Actor-Critic Methods};
% Second level nodes
\node[bluebox, below=1.5cm of value] (dp) {Dynamic Programming\\
Q-learning\\
SARSA};
\node[redbox, below=1.5cm of policy] (nonlinear) {
Policy Gradient Methods\\
Policy Search};
\node[purplebox, below=1.2cm of actorcritic] (actorcriticmethods) {A2C\\
SAC\\ DDPG};
% Third level nodes - aligned horizontally
\node[bluebox, below=1.2cm of dp] (ddqn) {DQN\\DDQN};
\node[redbox, below=1.2cm of nonlinear] (ppo_trpo) {TRPO\\
PPO};
% Arrows
\draw[arrow, tableaublue] (rl.south west) -- (value.north);
\draw[arrow, tableaugreen] (rl.south east) -- (policy.north);
\draw[arrow, tableaupurple] (rl.south) -- (actorcriticapprox.north);
\draw[arrow, tableaupurple] (actorcriticapprox.south) -- (actorcritic.north);
\draw[arrow, tableaublue] (value.south) -- (dp.north);
\draw[arrow, tableaugreen] (policy.south) -- (nonlinear.north);
\draw[arrow, tableaupurple] (actorcritic.south) -- (actorcriticmethods.north);
\draw[arrow, tableaublue] (dp.south) -- (ddqn.north);
\draw[arrow, tableaugreen] (nonlinear.south) -- (ppo_trpo.north);
\draw[arrow, tableaublue, dashed] (value.east) -- (actorcriticapprox.west);
\draw[arrow, tableaugreen, dashed] (policy.west) -- (actorcriticapprox.east);
\end{tikzpicture}
\caption{Reinforcement Learning Topology}
\label{fig:RL_topology}
\end{figure}

The key insight underlying model-free RL is that, while the transition dynamics $p(\bm{x}_{t+1} | \bm{x}_t, \bm{u}_t)$ may be unknown analytically, realizations of these transitions can be observed through system interaction. Model-free algorithms exploit this by learning to approximate optimal value functions or policies using only observed transition tuples $( \bm{x}_t, \bm{u}_t, \phi_t, \bm{x}_{t+1})$. These algorithms can be categorized based on their approximation strategy: value function approximation, direct policy approximation, or hybrid approaches combining both methodologies (Figure \ref{fig:RL_topology}). 

We begin by outlining some fundamentals of RL algorithm formulation in Section \ref{sec:6_1}, before describing value function approximation algorithms in Section \ref{sec:6_2} and policy approximation algorithms in Section \ref{sec:6_3}. 
Finally, Section \ref{sec:6_4} describes a popular class of hybrid approaches known as actor-critic algorthms.

%%%%%%%%%%%%%%%%%%%%%%%%

\subsection{Multi-Armed Bandits}\label{sec:6_1}
\subsubsection{Problem Formulation}

The bandit problem formulation was introduced in \cite{robbins1952some} and  represents a classical and simple case of sequential decision-making, where there exists no state space (Figure \ref{fig:bandit_mc_td}). 
Effectively, with no state space, taking any action $\bm{u}$ directly leads to a given reward realization. 
The agent repeatedly selects controls $\bm{u}_n$ and observes resulting costs $\phi_n$ over $N$ trials, aiming to minimize cumulative cost by identifying optimal controls.

In the bandit setting, a policy $\pi = (\pi_0, \pi_1, \ldots, \pi_{N-1})$ is a sequence of control selection rules, where $\pi_n(\bm{u}\mid \mathcal{D}_n)$ provides a probability distribution on controls conditional to the history of observations and chosen controls up to the current trial $n$, denoted $\mathcal{D}_n = (\bm{u}_0, \phi_0, \ldots, \bm{u}_{n-1}, \phi_{n-1})$. The objective of a bandit strategy is to solve
\begin{equation}\label{eq:bandit_problem}
   \min_{\pi}  \mathbb{E}_\pi\left[\sum_{n=0}^{N-1} \phi_n \mid \pi\right]\,,
\end{equation}
where the expectation is taken over the randomness in the realization of costs  $\phi \sim p(\phi|\bm{u})$, and the controls.

The central challenge lies in balancing exploration of the control space with exploitation of the currently best-performing controls---a concept that arises in many active learning settings often known as the \textit{exploration-exploitation tradeoff}. 
For instance, consider dual control, where identification and control are balanced by heuristically incorporating persistent excitation in a process control problem~\cite{mesbah2018stochastic}. 

Addressing this tradeoff for the bandit problem requires estimating control values $Q(\bm{u}) = \mathbb{E}[\phi \mid \bm{u}]$, through sample averaging $Q_n(\bm{u}) \approx Q(\bm{u})$, and defining policies that effectively manage the exploration-exploitation tradeoff. 
For simple distributions and reward models, computing and maximizing the Gittins index~\cite{gittins1979bandit} gives the optimal policy for solving \eqref{eq:bandit_problem}, but, like exact DP approaches, computing Gittins indices can quickly become computationally intractable. In PSE, we often encounter bandit-like problems in design-of-experiments and parameter estimation, where each experimental condition represents a control and the information gained serves as the cost signal. Crucially, these problems lack the state transitions present in RL formulations which is that the system's underlying parameters remain static, though our belief about them evolves. Bayesian optimization methods~\cite{paulson2025bayesian} exploit this structure by leveraging assumed smoothness in the objective function, enabling efficient sequential sampling policies without requiring a dynamics model.

\subsubsection{Bandit Algorithms}
Several established algorithms address the exploration-exploitation tradeoff in solving \eqref{eq:bandit_problem} in practice. Confidence bound approaches incorporate both estimated values and confidence intervals into a so-called upper/lower confidence bound (UCB/LCB, respectively)~\cite{auer2002finite}. 
Thompson Sampling~\cite{thompson1933likelihood} takes a Bayesian approach, maintaining posterior distributions over expected costs and sampling from them for control selection. 
Sampling from the posterior distribution provides an intuitive way to manage the exploration-exploitation tradeoff. 

The simple $\epsilon$-greedy policy prescribes a stationary policy where at each trial $n$, the control selection rule $\pi_n$ selects the greedy control with probability $(1-\epsilon)$ and explores randomly with probability $\epsilon$:
\begin{equation}\label{eq:epsilongreedy}
\pi_n(\mathcal{D}_n) = \begin{cases}
\arg\min_{\bm{u}} Q_n(\bm{u}) & \text{with probability } 1-\epsilon \\
\text{random control from } \mathcal{U} & \text{with probability } \epsilon.
\end{cases}
\end{equation}

This class of algorithms generally involves learning an approximation of the action-value estimates $Q(\bm{u})$. Note that in the bandit setting, the formulations from Section~\ref{sec:valuefunctions} are simplified, as the state space $\mathcal{X}$ is empty. 
Action-value estimates are updated incrementally using the formula:
\begin{equation*}
Q_{n+1}(\bm{u}) = Q_n(\bm{u}) + \frac{1}{N_n(\bm{u})}[\phi_n - Q_n(\bm{u})],
\end{equation*}
where $N_n(\bm{u})$ is the number of times control $\bm{u}$ has been selected up to trial $n$, and the step size $\frac{1}{N_n(\bm{u})}$ decreases as more samples of that specific control are collected, ensuring convergence to true expected costs.

\begin{figure}    
\centering
\begin{tikzpicture}[node distance=2.5cm,
    state/.style={
        circle,
        draw,
        minimum size=1.2cm,
        thick
    },
    action/.style={
        rectangle,
        rounded corners,
        draw,
        minimum size=1cm,
        thick
    },
    transition/.style={
        ->,
        >=Stealth,
        thick
    }
]
    % Bandit model (left side)
    \begin{scope}[xshift=2cm]
    \node[state] (s0) {$\bm{u}_n$};
    \node[state, below=1.5cm of s0] (a0) {$\phi_n$};
    \node[below right=1.15cm of s0, font=\Large] {$\ldots$};
    \draw[transition] (s0) -- node[left] {} (a0);
    \node[above=0.5cm of s0, font=\bfseries] {Bandit};
    \end{scope}
    % MDP model (right side)
    \begin{scope}[xshift=6cm, yshift=-1.5cm]
        \node[state] (s1) {$\bm{x}_t$};
        \node[state, right=2cm of s1] (s2) {$\bm{x}_{t+1}$};
        \node[state, below right=1cm of s1] (a1) {$\phi_t$};
        \node[state, above right=1cm of s1] (s3) {$\bm{u}_{t}$};
        
        \draw[transition] (s1) -- node[left] {} (a1);
        \draw[transition] (s2) -- node[below] {} (a1);
        \draw[transition] (s3) -- node[right] {} (s2);
        \draw[transition] (s1) -- node[above left] {} (s3);
        \draw[transition] (s3) -- node[right] {} (a1);
        
        \node[above=0.5cm of s3, font=\bfseries] {MDP};
        
        % Add ellipsis after MDP
        \node[right=0.2cm of s2, font=\Large] {$\ldots$};
    \end{scope}
    \end{tikzpicture}

    \caption{Directed probability graph for multi-armed bandits (left) and MDPs (right). The bandit probability graph shows the cost is only depenedent on the chosen arm where as the MDP graph shows the next state and cost incurred depends on both the control chosen, the current state and the next state.}
    \label{fig:bandit_mc_td}
    \end{figure}
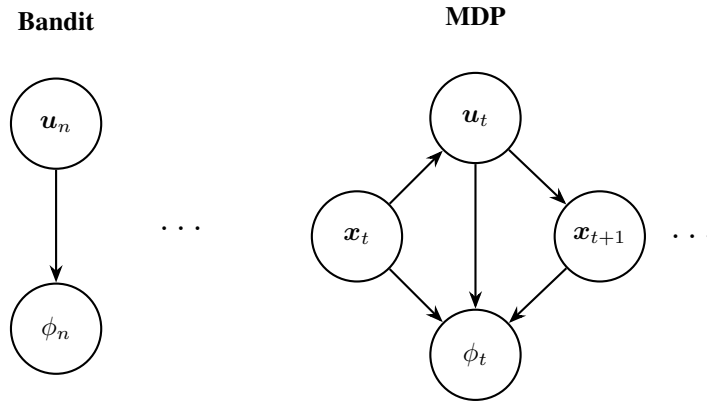

The connection between bandit problems (Equation \ref{eq:bandit_problem}) and Bayesian Optimization provides a valuable perspective for PSE design applications. 
Both frameworks balance exploration and exploitation in uncertain environments, with BO extending these concepts to continuous spaces through probabilistic surrogate models. This approach has proven effective in materials design~\cite{sorourifar2025adaptive, griffiths2020constrained}, reaction optimization~\cite{shields2021bayesian, gonzalez2023new, helleckes2024high}, and process engineering~\cite{savage_machine_2024, coutinho2023bayesian}. 
We refer the reader to \cite{paulson2025bayesian} for an overview of BO methods and applications in PSE. 
%AC Link comment
Importantly, the exploration strategies developed for bandits remain fundamental building blocks in the full RL setting, i.e., for MDPs. These data-collection strategies address the exploration-exploitation tradeoff that persists when a state space is present, i.e. when the problem is a contextual bandit, or when there is a decision process at hand (Figure~\ref{fig:bandit_mc_td}).  Indeed, balancing exploration and exploitation remains central to RL strategies such as state-action value learning and policy optimization. 

\subsection{State-action value learning}\label{sec:6_2}
State-action value learning describes a class of RL algorithms that are based on predicting future costs given potential state and control pairs (also referred to as state-action pairs). These methods focus on constructing a state-action value function (commonly called the Q-function), which returns a future cost for a given state-action pair. Being able to accurately predict state-action values allows the agent to form a policy which chooses the control that minimizes this function to retrieve the optimal policy as discussed in Section \ref{sec:exact_mdp}.

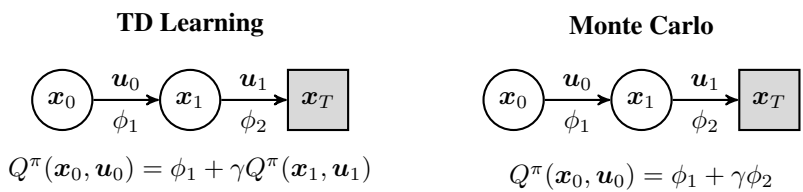
\begin{figure}
    \centering
\begin{tikzpicture}[
    node distance=2.5cm,
    state/.style={circle, draw, minimum size=0.8cm},
    terminal/.style={rectangle, draw, fill=gray!30, minimum size=0.8cm},
    thick,
    >=stealth',
    font=\sffamily
]
% Title
% Bandit Problems - single state with self-loops

% TD Learning
\begin{scope}[xshift=-3cm,yshift=2cm]
    \node[font=\bfseries] at (0,1) {TD Learning};
    
    \node[state] (S0) at (-1.7,0) {$\bm{x}_0$};
    \node[state] (S1) at (0,0) {$\bm{x}_1$};
    \node[terminal] (S2) at (1.7,0) {$\bm{x}_T$};
    
    % Actions added to TD Learning
    \draw[->] (S0) -- (S1) node[midway, above] {$\bm{u}_0$} node[midway, below] {$\phi_1$};
    \draw[->] (S1) -- (S2) node[midway, above] {$\bm{u}_1$} node[midway, below] {$\phi_2$};
    
 \node[below=0.2cm of S1] {$Q^\pi(\bm{x}_0,\bm{u}_0) = \phi_1 + \gamma Q^\pi(\bm{x}_{1}, \bm{u}_{1})$};
\end{scope}
% MC Learning
\begin{scope}[xshift=3cm, yshift=2cm]
    \node[font=\bfseries] at (0,1) {Monte Carlo};
    
    \node[state] (S0) at (-1.7,0) {$\bm{x}_0$};
    \node[state] (S1) at (0,0) {$\bm{x}_1$};
    \node[terminal] (S2) at (1.7,0) {$\bm{x}_T$};
    
    % Actions added to Monte Carlo
    \draw[->] (S0) -- (S1) node[midway, above] {$\bm{u}_0$} node[midway, below] {$\phi_1$};
    \draw[->] (S1) -- (S2) node[midway, above] {$\bm{u}_1$} node[midway, below] {$\phi_2$};
    
    % Return G at terminal state
    \node[below=0.3cm of S1] {$Q^\pi(\bm{x}_0,\bm{u}_0) = \phi_1 + \gamma \phi_2$};
\end{scope}
\end{tikzpicture}
\caption{Comparison of RL paradigms: temporal difference learning (left) and Monte Carlo methods (right). Temporal difference methods use a single transiition cost and the Q-function to estimate the state-action value whereas Monte Carlo methods use the entire trajectory of costs.}
\label{fig:TD_MC}
\end{figure}

\subsubsection{Monte Carlo Learning}

Monte Carlo (MC) methods in RL represent an approach for solving sequential decision problems without knowledge of the underlying system dynamics (namely model-free RL) \cite{sutton1998reinforcement}. Assuming a finite horizon, MC methods operate in environments where controls influence both immediate costs and future states, and that terminate after a finite number of steps \cite{benyahia2021control,puterman2014markov}.

The defining characteristic of MC methods is their reliance on complete episodes of experience to learn, necessitating the finite-horizon assumption~\cite{rubinstein2016simulation,curtiss1954theoretical}. This characteristic can lend itself useful for the batch chemical processes, which naturally have a finite horizon, with performance often only measured at the end-state. Given a finite horizon $\mathcal{T}\setminus\{T\}$, an episode represents a complete trajectory $\tau = (\bm{x}_0, \bm{u}_0, \phi_0, \bm{x}_1, \bm{u}_1, \phi_1, \ldots, \bm{x}_{T-1}, \phi_{T-1})$ from an initial state to a terminal state following a specific policy $\bm{u}_t \sim \pi_t(\cdot|\bm{x}_t)$. Rather than updating estimates after each individual transition, as in methods discussed later in the review, MC methods wait until an episode concludes, then use the observed sum of the cost trajectory to update value estimates for every state-action pair encountered (see Figure~\ref{fig:TD_MC}). 
In this way, they avoid having to apply the Bellman equation~\ref{eq:bellman_optimality} to produce value function estimates. 

Specifically, for a state-action pair $(\bm{x}_t, \bm{u}_t)$ visited at time $t$ during a full trajectory (episode), the return $G_t$ is defined as the sum of discounted future costs from that time step onward,
\begin{equation}\label{eq:return}
G_t = \sum_{k=t}^{T-1} \gamma^{k-t} \phi_k .
\end{equation}
This return provides a sample that can be used to learn the true action-value function $Q^{\pi}(\bm{x}_t, \bm{u}_t) = \mathbb{E}_{\pi}[G_t | \bm{x}_t, \bm{u}_t]$, which represents the expected future return when taking control $\bm{u}_t$ in state $\bm{x}_t$ and following policy $\pi$ thereafter. 
Note that samples are collected following the current policy $\pi$, which is normally not the optimal policy $\pi^*$. 
After each episode, the state-action values can be updated incrementally,
\begin{equation*}
Q^{\pi}(\bm{x}_t, \bm{u}_t) \leftarrow Q^{\pi}(\bm{x}_t, \bm{u}_t) + \alpha [G_t - Q^{\pi}(\bm{x}_t, \bm{u}_t)],
\end{equation*}
where $\alpha \in (0,1]$ is the learning rate that controls the step size of the updates, balancing the influence of new returns against accumulated knowledge. As the number of episodes increases, these sample-based estimates converge to the true expected values under policy $\pi$.

Noting the dependence of the learned value functions on the applied policy $\pi$, MC methods are typically classified based on the relationship between the policy used to generate data and the policy being evaluated or improved. The simplest approach is on-policy learning, where the same policy is used for both exploration and evaluation~\cite{singh1996reinforcement}. In on-policy MC, the agent follows policy $\pi$ to generate episodes, and the resulting data is used to estimate $Q^{\pi}$. The policy is then improved, e.g., by defining the next policy to greedily maximize the current Q-function estimates as shown by Equation~\eqref{eq:opt_pol_q}~\cite{sutton1998reinforcement}.

On-policy methods face a dilemma: they seek to learn control values conditional on subsequent optimal behavior, but they need to behave non-optimally in order to explore all controls to find the optimal policy. 
In effect, this is the exploration-exploitation tradeoff again, from a slightly different lens. 
Therefore the on-policy approach must learn control values not for the current guess of the optimal policy, but for a near-optimal policy that incorporates exploration. For value-based on-policy methods, this is typically implemented by modifying the policy to be $\epsilon$-greedy, selecting the greedy control with probability $(1-\epsilon)$ and a random control with probability $\epsilon$, analogous to \eqref{eq:epsilongreedy}. The exploration-exploitation balance can be controlled by gradually decreasing $\epsilon$ as learning progresses. 

On the other hand, off-policy MC methods address the exploration-exploitation tradeoff by explicitly decoupling the behavior policy used for exploration from the target policy being evaluated or improved such as by using importance sampling. This approach allows the agent to learn from data generated by any policy, including historical process data or simulation models with different control strategies. For the interested reader we refer to ~\cite{sutton1998reinforcement}.

% To enable off-policy learning, MC methods employ importance sampling to correct for the distribution mismatch between policies.  In the off-policy setting, the target policy $\pi$ represents the policy being evaluated or improved, while the behavior policy $\mu$ denotes the policy that generated the episode data. The behavior policy $\mu$ must satisfy the coverage condition $\mu(\bm{u}_t\mid\bm{x}_t) > 0$ whenever $\pi(\bm{u}_t\mid\bm{x}_t) > 0$ to ensure that all controls potentially selected by the target policy have non-zero probability under the behavior policy. The importance sampling ratio for a trajectory generated by the behavior policy is defined as,
% \begin{equation}\label{eq:importance_sampling}
% \rho_{t} = \prod_{k=t}^{T-1} \frac{\pi(\bm{u}_k|\bm{x}_k)}{\mu(\bm{u}_k|\bm{x}_k)} .
% \end{equation}
% This ratio weights the returns \eqref{eq:return} when updating the Q-function,
% \begin{equation*}
% Q^{\pi}(\bm{x}_t, \bm{u}_t) \leftarrow Q^{\pi}(\bm{x}_t, \bm{u}_t) + \alpha [\rho_{t} G_t - Q^{\pi}(\bm{x}_t, \bm{u}_t)] .
% \end{equation*}
% While importance sampling \eqref{eq:importance_sampling} provides unbiased estimates, it introduces high variance for long trajectories as the probability ratios compound~\cite{precup2000eligibility, szepesvari2022algorithms}. This variance challenge is particularly problematic in process control applications where reliability and consistency are essential. 
The episodic nature of batch operations aligns well with the requirement of MC-learning for complete episodes. For example, recent work has shown that MC-based RL outperforms TD-learning for batch polymerization processes, providing more stable policy updates where early controls have irreversible consequences on final product quality~\cite{YOO2021107133}. Their approach showed superior constraint satisfaction when controlling batch polyol production under parametric uncertainty, leveraging MC methods' ability to account for complete trajectory returns and critical end-point constraints. 

 Despite these advantages for episodic batch operations, MC methods remain underexplored in the broader PSE literature. This may be because their requirement for complete episodic samples means that they are ultimately impractical for continuous processes with extended (or infinite) operating horizons. In general, the literature suggests preference of TD-learning over MC-learning from practitioners in PSE.

\subsubsection{SARSA}

Temporal Difference (TD) learning represents an alternative to MC methods, where agents can learn from individual transitions rather than requiring complete episodes. SARSA (State-Action-Reward-State-Action) is an on-policy TD learning algorithm that updates the Q-function estimate, i.e., a learned estimate of $Q^*(\bm{x},\bm{u})$ based on observed state transitions and costs~\cite{SARSA}.

The central concept in TD-learning is the temporal difference error, defined as the difference between the observed cost plus discounted estimated future value and the current estimate,
\begin{equation*}
\delta_t = \phi_{t} + \gamma Q^\pi(\bm{x}_{t+1}, \bm{u}_{t+1}) - Q^\pi(\bm{x}_t, \bm{u}_t).
\end{equation*}
This formulation mirrors the Bellman expectation operator structure from \eqref{eq:bellman_expectation} however, uses a single sample transition rather than the full expectation.
This TD error $\delta_t$ quantifies the prediction error in the current estimate and serves as the direction for value function updates, with the learning rate controlling the step size of these corrections. This online, error-driven update structure parallels adaptive control~\cite{annaswamy2023adaptive}, where prediction errors drive continuous parameter adjustments without requiring complete trajectory information. Unlike MC methods, which wait until the end of an episode to update value estimates, SARSA performs updates after each transition using the observed tuple $( \bm{x}_t, \bm{u}_t, \phi_{t}, \bm{x}_{t+1}, \bm{u}_{t+1} )$. 
The SARSA algorithm then updates the state-action value function for the current state,
\begin{equation*}
Q^{\pi}(\bm{x}_t, \bm{u}_t) \leftarrow Q^{\pi}(\bm{x}_t, \bm{u}_t) + \alpha \delta_t.
\end{equation*}

This update uses the current Q-value estimate of the next state-action pair, weighted by the discount factor $\gamma$, rather than waiting for the complete return. SARSA is classified as an on-policy algorithm, because it evaluates and improves the same policy used to generate the data. By learning the value of the actual policy being followed (including exploratory controls), SARSA will learn an exploratory and, therefore, suboptimal policy~\cite{sutton1998reinforcement}. Perhaps owing to this limitation of SARSA, there is a lack of work from the PSE community using this algorithm; however, it forms the basis for introducing the popular Q-learning algorithms, described in the following section.

\subsubsection{Q-learning}
% Off-policy TD learning
% Limitations in state space
% breifly TD(lambda)/n-step to bridge between mc 

Q-learning is an off-policy TD method that, per its name, directly approximates the optimal action-value function (Q function), regardless of the policy being followed \cite{watkins_q-learning_1992}. Unlike SARSA, which learns the action-value function subject to the policy being executed (including exploratory controls), Q-learning seeks to learn the value of the optimal policy, while potentially following a different exploratory policy. This decoupling of learning and behavior makes Q-learning especially valuable.
The core update rule for Q-learning uses the observed tuple $( \bm{x}_t, \bm{u}_t, \phi_{t}, \bm{x}_{t+1} )$ to update the state-action value function,
\begin{equation}\label{eq:qlearning_update}
Q^{\pi}(\bm{x}_t, \bm{u}_t) \leftarrow Q^{\pi}(\bm{x}_t, \bm{u}_t) + \alpha[\phi_{t} + \gamma \min_{\bm{u}_{t+1}} Q^{\pi}(\bm{x}_{t+1}, \bm{u}_{t+1}) - Q^{\pi}(\bm{x}_t, \bm{u}_t)].
\end{equation}

The key distinction from SARSA lies in the target term, which combines the observed cost with the optimal state value (expected cost-to-go) from the next state. This assumes that from the next state, one will select the optimal control, irrespective of the control $\bm{u}_{t+1}$ that might actually be taken. This decoupling of the target and behavior policies allows Q-learning to learn about the optimal policy while following a suboptimal (exploratory) policy. The theoretical guarantee of Q-learning is that, given sufficient exploration and under appropriate learning rate schedules, the Q-function will converge to the optimal action-value function $Q^*$ with probability 1 ~\cite{watkins1992qlearningproof}.
% Addresses AC comment about transition to next section using the Q-table
In practice, tabular Q-learning maintains a table of Q-values for each discrete state-action pair. At each time step, the agent selects a control, with the behavioural policy, observes the resulting state and cost, then updates the corresponding Q-table entry using the update rule given in \eqref{eq:qlearning_update}. The policy can be derived at any time by selecting the action with the minimum Q-value for each state. This tabular approach works well for problems with small, discrete state and control spaces, but becomes computationally intractable for sets with high cardinality or the continuous spaces commonly encountered in PSE applications.

\subsubsection{Deep Q-Network}
\begin{figure}
    \centering
    \includegraphics[width=1\linewidth]{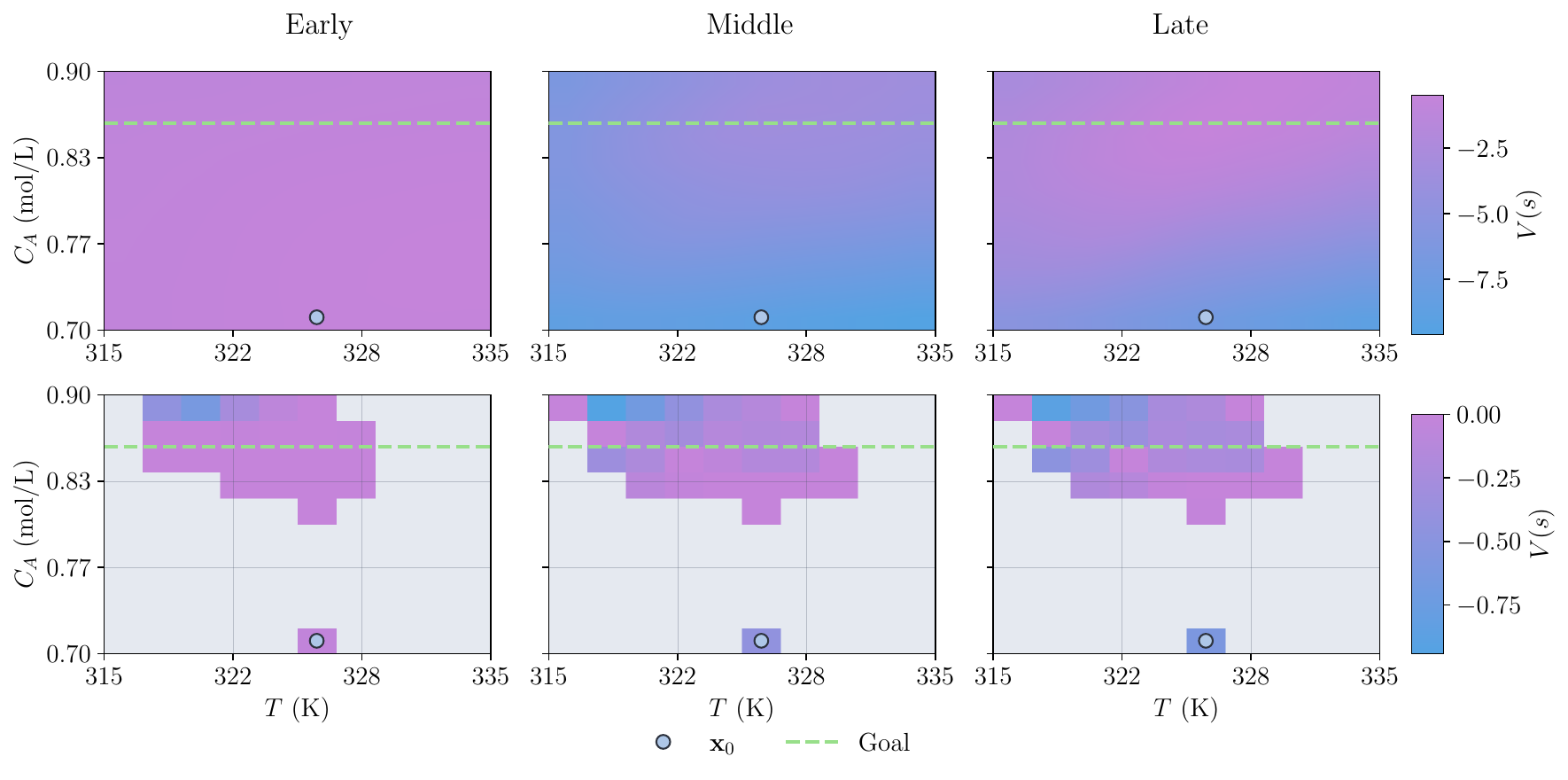}
    \caption{Comparision of Q-learning (bottom row) and DQN (top row) stages of learning. The value function is shown with a heatmap with pink being high value and blue being low value. DQN defines a value function over the entire \textit{continuous} state space whereas Q-learning maintains a table of values for visited states.}
    \label{fig:DQN_Q_learning}
\end{figure}

High cardinality or continuous states spaces pose a significant challenge for tabular state-action value learning algorithms such as Q-learning, which require storing a value for each discrete state-action pair (illustrated for states in the bottom row of Figure~\ref{fig:DQN_Q_learning}). When the state space becomes continuous (as depicted in the top row of Figure~\ref{fig:DQN_Q_learning}) or has high cardinality, explicit enumeration/discretization and storage can quickly become computationally infeasible. To address this limitation, the Q-function can be approximated by a parameterized function and learned. This idea was exemplified by the algorithm Deep Q-Networks (DQN) \cite{mnih2013playingatarideepreinforcement}, which utilizes deep neural networks to parameterize the action-value function with parameters $\bm{\omega}\in \mathbb{R}^{n_\omega}$, denoted as $Q_\omega(\bm{x}_t, \bm{u}_t)$.

The optimization objective for training such a parameterized function minimizes the expected squared error between predicted Q-values and target values derived from the Bellman equation,
\begin{equation}\label{eqn:dqn_loss}
L(\omega) = \mathbb{E}_{ ( \bm{x}_t,\bm{u}_t,\bm{x}_{t+1},\phi_t) \sim\mathcal{D}} \left[ \left( \phi_t + \gamma \min_{\bm{u}_{t+1}} Q^{\pi}_{\omega'}(\bm{x}_{t+1}, \bm{u}_{t+1}) - Q^{\pi}_\omega(\bm{x}_t, \bm{u}_t) \right)^2 \right],
\end{equation}
where $\mathcal{D}$ represents a dataset of transition tuples and $\omega'\in \mathbb{R}^{n_\omega}$ denotes parameters of a second, target network. Two key mechanisms stabilize DQN learning: (1) experience replay, which maintains a so-called `replay buffer' of past transitions, from which mini-batches are randomly sampled, and (2) as alluded to previously a target network with parameters $\omega'$ that are periodically updated to match the primary network. These mechanisms break temporal correlations in sequential data and stabilize learning targets, addressing instabilities inherent in nonlinear function approximation. The second point ultimately means that gradients $\nabla_\omega L(\omega)$ derived from this objective are semi-gradients (approximate gradients) for the actual squared Bellman error.

Once the Q-function approximation is trained, the DQN selects actions by evaluating the Q-function for all possible controls and choosing the one with minimum predicted cost. This action selection procedure inherently requires a discrete (or discretized) control space, as it necessitates explicit enumeration and evaluation of all possible actions. In practical implementations, most DQN algorithms employ a model architecture with multiple outputs (one for each possible control), therefore giving the state-action values for all possible actions in a single model evaluation~\cite{mnih2013playingatarideepreinforcement}. 
Selecting a control then reduces to identifying the minimal model output. 
For continuous control spaces, controls can be selected by optimizing over the trained neural network~\cite{BURTEA2024108518, ryu2019caql}, i.e., solving for the inputs that minimize the state-action value function. This generally requires solving a nonlinear and nonconvex optimization problem. This can be implemented  via local solvers, e.g. interior point methods \cite{wright1997primal}, provided the formulation is defined on continuous decision variables (i.e., do not rely on mixed-integer reformulations, as one may well require if the Q-network is defined as a piece-wise affine function) and if a local solution is acceptable. 
% PSE works that use dqn

DQN's applications in PSE have demonstrated promising results. For example, \citet{HWANGBO2020106910} applied DQN to control liquid-liquid extraction columns in downstream biopharmaceutical processing, discretizing the continuous control domain into 35 distinct controls to avoid the additional optimization of the q-network. Their implementation integrated Monte-Carlo sampling to handle process uncertainties, achieving an improvement in API recovery yield compared to open-loop operation. Similarly, ~\citet{konishi2022fluid} employed DQN for optimizing fluid mixing operations, where the algorithm learned effective flow control strategies by discretizing the flow parameter space. Their implementation achieved exponentially fast mixing without prior system knowledge and established an efficient transfer learning methodology between different fluid regimes. While these studies demonstrate DQN's capacity to handle complex nonlinear processes through appropriate discretization strategies, they simultaneously highlight a fundamental limitation of value-based methods: their inability to directly handle continuous control spaces prevalent in PSE applications. While one approach to extend DQN to continuous control involves directly optimizing over the Q-function to select actions~\cite{BURTEA2024108518, ryu2019caql}, this can introduce significant computational challenges, especially if a global solution is required. This limitation has motivated the development of policy-based and actor-critic methods that can offer more tractable solutions for continuous control problems.
\subsection{Policy Optimization} \label{sec:6_3}
% Find the optimal policy without using the Q-function
Policy optimization methods represent an alternative class of algorithms for solving the RL problem compared to the value-based methods discussed above. Instead of explicitly modeling the value function and deriving a policy from it, policy optimization methods work directly in the policy space and optimize the policy itself. These methods can either compute gradient estimates to perform gradient descent in the policy space or employ derivative-free approaches that directly search the parameter space to minimize expected cost.

\subsubsection{Policy Gradient}\label{sec:policy_gradient}

Policy gradient methods provide a strong theoretical foundation for direct policy optimization~\cite{PG_theorem}. These approaches parameterize the policy $\pi_\theta(\bm{u}_t\mid\bm{x}_t)$ with parameters $\theta \in \mathbb{R}^{n_\theta}$, then optimize these parameters to minimize the expected cost~\cite{REINFORCE}. The fundamental objective in policy gradient methods is to minimize the expected cumulative discounted cost,
\begin{equation}\label{eq:pg_obj}
J(\theta) = \mathbb{E}_{p_{\pi_\theta}}\left[\sum_{t=0}^{T-1} \gamma^t \phi_t\right],
\end{equation}
where $p_{\pi_\theta}$ represents the distribution over trajectories $\bm{\tau}$ induced by following a fixed policy $\pi_\theta$:
\begin{equation}
 p_{\pi_\theta}(\bm{\tau}) = p_0(\bm{x}_0) \prod_{t=0}^{T-1} \pi_\theta(\bm{u}_t|\bm{x}_t) p(\bm{x}_{t+1}|\bm{x}_t,\bm{u}_t).
\end{equation}
To derive the policy gradient, the gradient of $J(\theta)$ is taken and the log-derivative trick is applied\footnote{Log-derivative trick: $\nabla_\theta p_{\pi_\theta}(\bm{\tau}) = p_{\pi_\theta}(\bm{\tau}) \nabla_\theta \log p_{\pi_\theta}(\bm{\tau})$}
\begin{equation}
\nabla_\theta J(\theta) = \mathbb{E}_{p_{\pi_\theta}}\left[\nabla_\theta \log p_{\pi_\theta}(\bm{\tau}) \sum_{t=0}^{T-1} \gamma^t \phi_t\right].
\end{equation}
Since the dynamics $p(\bm{x}_{t+1}|\bm{x}_t, \bm{u}_t)$ are independent of $\theta$, only the policy terms contribute to the inside term,
\begin{equation}
\nabla_\theta \log p_{\pi_\theta}(\bm{\tau}) = \sum_{t=0}^{T-1} \nabla_\theta \log \pi_\theta(\bm{u}_t|\bm{x}_t).
\end{equation}
This is known as a semi-gradient where only the policy is differentiated, excluding the dynamics or costs, avoiding the need to explicitly compute or estimate the gradients related to the environment.
Since only the policy terms in $\log p_{\pi_\theta}(\bm{\tau})$ depend on $\theta$, we have $\nabla_\theta \log p_{\pi_\theta}(\bm{\tau}) = \sum_{t=0}^{T-1} \nabla_\theta \log \pi_\theta(\bm{u}_t|\bm{x}_t)$. Applying causality (actions at time $t$ only affect costs from time $t$ onward), we obtain
\begin{equation}
\nabla_\theta J(\theta) = \mathbb{E}_{p_{\pi_\theta}}\left[\sum_{t=0}^{T-1} \nabla_\theta \log \pi_\theta(\bm{u}_t|\bm{x}_t) G_t\right],
\end{equation}
where $G_t = \sum_{k=t}^{T-1} \gamma^{k-t} \phi_{k}$ is the discounted return from time step $t$. This expression, known as the policy gradient \cite{PG_theorem}, is central to a number of different algorithms despite being an approximation to the gradient of \eqref{eq:pg_obj}.
The REINFORCE algorithm \cite{REINFORCE} directly estimates the policy gradient by sampling trajectories and then performs updates,
\begin{equation*}
\theta \leftarrow \theta - \alpha \sum_{t=0}^{T-1} \nabla_\theta \log \pi_\theta(\bm{u}_t|\bm{x}_t)  G_t,
\end{equation*}
where $\alpha$ is the learning rate. This approach suffers from high variance in the gradient estimates, potentially leading to unstable learning.

To address the high variance that leads to unstable learning, a state-dependent baseline can be introduced, which modifies the policy gradient update. Since $b(\bm{x}_t)$ depends only on the state and not the action, it can be factored out of the expectation over $\bm{u}_t$,
$$\mathbb{E}_{p_{\pi_\theta}}[ \nabla_\theta\log\pi_\theta(\bm{u}_t \mid \bm{x}_t)b(\bm{x}_t)] = \mathbb{E}_{\bm{x}_t}\left[b(\bm{x}_t) \mathbb{E}_{\bm{u}_t \sim \pi_\theta(\cdot|\bm{x}_t)}[\nabla_\theta\log\pi_\theta(\bm{u}_t|\bm{x}_t)]\right] = 0,$$
where the inner expectation equals zero by the log probability lemma. This is also commonly known as a control variate which may be more familiar to the PSE community~\cite{giles2015multilevel}. While any state-dependent baseline leaves the gradient unbiased, the choice of baseline significantly affects variance. The on-policy value function $V^{\pi_\theta}(\bm{x}_t) = \mathbb{E}_{p_{\pi_\theta}}[G_t \mid \bm{x}_t]$ is particularly effective at reducing variance, as it approximates the expected return from each state. Since this function also has to be approximated, it leads to a new class of algorithms called actor-critic algorithms, which are discussed in Section \ref{sec:6_4}.

Policy gradient methods have demonstrated success in PSE applications, most notably in batch bioprocess optimization under uncertainty. \citet{petsagkourakis2020reinforcement} developed a framework using the REINFORCE algorithm~\cite{REINFORCE} with recurrent neural networks (RNNs) to parameterize control policies for bioprocesses. RNNs were selected for their ability to capture temporal dependencies in sequential decision problems, with the network outputting parameters of a Gaussian distribution from which control actions were sampled. Their two-phase approach (initial training on an approximate model followed by transfer learning for batch-to-batch adaptation) minimized required sampling from the actual process. The method outperformed nonlinear model predictive control across three case studies involving stochastic differential equations and nonsmooth dynamics, while demanding substantially less online computation. This work demonstrates how policy gradients can effectively handle complex chemical processes where traditional optimization approaches struggle with stochasticity and nonlinearities.

\subsubsection{Policy Search}
% Challenges for stochastic approximation
While policy gradient methods rely on stochastic gradient estimation through trajectory sampling and analytical gradients, policy search methods approach policy optimization as a black-box optimization problem. Rather than computing gradients of the parameterized policy, these methods directly search the parameter space to maximize returns, often without requiring differentiability of the cost function or policy; random search has even been shown to be competitive for learning good policies in RL applications~\cite{benjamin2018random}.

In mathematical terms, policy search algorithms directly optimize the policy parameters $\theta$ by treating the expected return as a black-box function to be maximized,
\begin{equation*}
    \theta^* \in \arg\min_{\theta} J(\theta).
\end{equation*}
Policy search algorithms utilize various derivative-free optimization techniques to explore the parameter space. One common approach is to use evolutionary strategies, which have been shown to be a scalable alternative to gradient-based RL~\cite{salimans2017evolution}. These strategies maintain a population of candidate solutions (policy parameters) and iteratively improve them through selection, recombination, and mutation operations. 
We refer the reader to \cite{bartz2014evolutionary} for an overview of the underlying evolutionary algorithms.

Furthermore, a significant area of research focuses on creating hybrid algorithms that combine the strengths of both evolutionary and gradient-based approaches. The goal is to hybridize the broad, global search capabilities of evolutionary methods with the efficient local convergence of gradients. Several successful frameworks have emerged from this line of work, demonstrating the power of combining the two areas \cite{khadka2018evolutionguidedpg, pourchot2019cemrlcombiningevolutionarygradientbased, Bodnar_2020}.

Several PSE applications have leveraged policy search approaches. For instance, \citet{bloor2025cirl} introduced Control-Informed RL (CIRL), which embeds a PID controller structure within a deep RL policy network. They used evolutionary strategies to optimize the parameters of this hybrid policy for controlling a CSTR, finding it effective for tuning the combined neural network and non-differentiable PID components. \citet{qiu2024leveraging} applied multi-objective evolutionary algorithms (MOEAs), a population-based policy search technique, to find a diverse set of RL policies (a Pareto front) for dynamic decision-making in supply chain management under conflicting objectives. The derivative-free nature of MOEAs was crucial for handling the multi-objective and potentially non-differentiable nature of the problem. \citet{wu2023distributional} also tackled inventory management in supply chains, employing a hybrid derivative-free policy search algorithm combining ES, Artificial Bee Colony (ABC), PSO, and Simulated Annealing (SA). This approach demonstrated superior sample efficiency and performance compared to standard policy gradient methods for their problem; this work also explored distributional RL concepts (discussed further in Section \ref{sec:distributional_rl}). Policy search algorithms have proven themselves as a practical approach as they do not rely on potentially noisy, approximate gradient estimates as seen in policy gradient methods. Moreover they inherit the ability of global search methods to escape poor performing regions of the policy parameter space. However, they remain relatively underexplored considering the number works using actor-critic algorithms. For a broader survey on policy search methods we direct the reader to~\citet{sigaud2023combining}. 

\subsection{Actor-Critic}\label{sec:6_4}
% Parameterise both policy and q-function. Be careful with actor/critic notation and f
Actor-critic methods are a popular class of hybrid RL algorithms, integrating elements from both value-based and policy-based approaches~\cite{sutton1998reinforcement}. These methods employ two distinct parameterized components: an \textit{actor}, which represents the policy function dictating control selection, and a \textit{critic}, which represents a learned value function used to evaluate the actor's policy and guide its improvement~\cite{konda1999actor}. As such, these algorithms involve learning both a policy (the actor) and a value function (the critic). The motivation behind this class of algorithms is the reduction of variance of policy gradient~\cite{hafner2011reinforcement} methods while maintaining ability to handle continuous control spaces, which is vital in PSE applications. Moreover, by training an actor to approximate the computationally expensive Q-function maximization, these methods achieve greater computational efficiency during policy execution while preserving the benefits of value-based guidance for policy improvement.

Formally, the actor parameterizes the policy $\pi_{\theta}(\bm{u}_t|\bm{x}_t)$ using parameters $\theta \in \mathbb{R}^{n_\theta}$. The critic approximates either the state-value function $V_{\psi}(\bm{x})$ with parameters $\psi \in \mathbb{R}^{n_\psi}$, or the action-value function $Q_{\omega}(\bm{x}, \bm{u})$ with parameters $\omega \in \mathbb{R}^{n_\omega}$.

Conceptually, actor-critic methods can be viewed as an extension of policy gradient methods~\cite{REINFORCE}. Standard policy gradient approaches, as discussed in Section~\ref{sec:policy_gradient}, often suffer from high variance in the gradient estimates derived from MC returns~\cite{szepesvari2022algorithms}, potentially leading to unstable convergence. The introduction of a critic addresses this limitation by providing a learned value function estimate that serves as an estimate of the discounted return, thereby mitigating variance, but introducing bias into the gradient estimates.

Within the actor-critic framework, the policy gradient update is modified to leverage the critic's value estimates. When employing an action-value function critic, $Q_{\omega}(\bm{x}, \bm{u})$, the policy gradient $\nabla_{\theta} J(\theta)$ can be approximated as,
\begin{equation}
\nabla_{\theta} J(\theta) \approx \mathbb{E}_{p_{\pi_{\theta}}} \left[ \nabla_{\theta} \log \pi_{\theta}(\bm{u}_t|\bm{x}_t) Q^\pi_{\omega}(\bm{x}_t, \bm{u}_t) \right].
\label{eq:ac_policy_gradient}
\end{equation}
Here, $Q^\pi_{\omega}(\bm{x}_t, \bm{u}_t)$ estimates the expected cumulative future cost incurred upon taking control $\bm{u}_t$ in state $\bm{x}_t$ and subsequently adhering to policy $\pi$. The critic's parameters, $\omega$ (or $\psi$ if using $V^\pi_{\psi}$), are typically updated using temporal difference (TD) learning methods, e.g., by minimizing the expected squared TD error as in \eqref{eqn:dqn_loss}.

To further enhance stability and potentially accelerate learning, many actor-critic algorithms utilize an advantage function~\cite{schulman2015high}, $A(\bm{x}_t, \bm{u}_t)$. The advantage function quantifies the relative merit of taking control $\bm{u}_t$ in state $\bm{x}_t$ compared to the expected value of controls in that state according to the current policy. It is formally defined as:
\begin{equation*}
A^{\pi}_{\omega, \psi}(\bm{x}_t, \bm{u}_t) = Q^{\pi}_{\omega}(\bm{x}_t, \bm{u}_t) - V^{\pi}_{\psi}(\bm{x}_t).
\end{equation*}
In this formulation, the state-value function $V_{\psi}(\bm{x}_t)$ acts as a baseline, analogous to baseline subtraction techniques in standard policy gradient methods used to reduce variance in the semi-gradient estimate \cite{REINFORCE, petsagkourakis2020reinforcement}. Employing the advantage function $A_{\omega, \psi}(\bm{x}_t, \bm{u}_t)$ in place of $Q_{\omega}(\bm{x}_t, \bm{u}_t)$ within the policy gradient expression (\ref{eq:ac_policy_gradient}) typically leads to a significant reduction in semi-gradient variance.

A variety of actor-critic algorithms have been proposed, differing primarily in their specific implementations of the actor and critic updates, variance reduction techniques, and mechanisms for ensuring stable learning. Notable examples include:
\begin{itemize}
    \item \textbf{Advantage Actor-Critic (A2C)}~\cite{A2C}, which directly uses the TD error as an estimate of the advantage function.
    \item \textbf{Trust Region Policy Optimization (TRPO)}~\cite{TRPO} and \textbf{Proximal Policy Optimization (PPO)}~\cite{ppo}, which impose constraints on the policy updates to prevent large, destabilizing changes.
    \item \textbf{Deep Deterministic Policy Gradient (DDPG)}~\cite{DDPG}, which adapts the actor-critic framework for continuous control spaces using a deterministic actor policy.
    \item \textbf{Soft Actor-Critic (SAC)}~\cite{SAC}, which introduces entropy maximization into the objective function, controlled by a temperature parameter, to encourage exploration and improve robustness.
    \item \textbf{Twin Delayed DDPG (TD3)}~\cite{TD3}, which mitigates Q-value overestimation bias, a common issue in DDPG, through the use of clipped double Q-learning and delayed policy updates.
\end{itemize}
These algorithms employ various implementation strategies to address the challenges of deep RL, including target networks for Q-function stabilization~\cite{mnih2013playingatarideepreinforcement}, replay buffers for decorrelated updates~\cite{mnih2013playingatarideepreinforcement}, and gradient clipping for training stability~\cite{pascanu2013difficulty}. Advanced implementations further incorporate prioritized experience replay~\cite{schaul2015prioritized} and multi-step returns to improve learning efficiency and reduce bias-variance tradeoffs~\cite{mnih2016asynchronous}.

Actor-critic methods are prevalent in contemporary RL, particularly for addressing complex, continuous control problems. Their ability to balance sample efficiency with effective handling of continuous state and control spaces makes them highly suitable for PSE applications. Consequently, they represent a dominant algorithmic class in recent PSE studies involving dynamic control challenges, as illustrated by the examples summarized in Table \ref{tab:rl_overview}.

\begin{table}[htbp]
\centering
\caption{Overview of Actor-Critic Applications in PSE}
\label{tab:rl_overview}
\begin{tabular}{@{} c c c @{}}
\toprule
\textbf{Reference} & \textbf{Method} & \textbf{Application} \\
\midrule
\citet{dogru2021online} & A2C & Hybrid tank system \\
\citet{lawrence2022deep} & TD3 & Two-Tank system \\
\citet{patel2023practical} & DDPG & Distillation Column \\
\citet{ma2019continuous} & DDPG & Polymerization Reactor\\
\citet{dutta2023multiple} & DDPG & Multiple reactor case studies \\
\citet{bangi2021deep} & DDPG & Hydraulic fracturing \\
\citet{ballard2024reinforcement} & TD3 & Polymerization reactor \\

\bottomrule
\end{tabular}
\end{table}

While these applications demonstrate the versatility of actor-critic methods, high-dimensional PSE systems present tractability challenges that require additional considerations. Hierarchical frameworks, where high-level policies set goals for low-level controllers, offer one pathway for managing complexity~\cite{bloor2025hierarchicalrlmpcdemandresponse}. Multi-agent RL, where distributed agents coordinate local objectives, provides another approach particularly relevant for supply chains~\cite{kotecha2025leveraging} and multi-unit plants~\cite{yifei2023multi}. Additionally, state space reduction through domain knowledge (selecting relevant state variables rather than monitoring all process measurements) could improve sample efficiency. These strategies enable actor-critic methods to scale beyond the benchmark problems shown in Table~\ref{tab:rl_overview} to the large-scale systems common in industrial PSE applications.

%%%%%%%%%%%%%%%%%%%%%%%%%%%%%%%%%
\section{Distributional RL}\label{sec:distributional_rl}

The RL methods discussed operate using the expected value of returns as the objective, describing the distribution of costs observed under a policy to a point estimate of its first moment (see Figure~\ref{fig:dist_rl}). While computationally convenient, this simplification discards valuable information about risk, uncertainty, and the multimodal nature of process outcomes. 
This is analogous to stochastic programming, where expected-value objectives are often preferred to risk-aware formulations. 
Distributional RL~\cite{bellemare2017distributional, bellemare2023distributional} addresses this limitation by modeling the entire distribution of returns, or at least some higher-order aspects (variance, tails, etc.), rather than merely their expected values.

\subsection{Distributional Framework}

In the standard RL formulations, such as those described in Section~\ref{sec:6_2}, the state-action value function models the expected return. Distributional RL extends this concept by considering the random return,
\begin{equation}\label{eq:randomreturn}
Z^\pi(\bm{x}_t, \bm{u}_t) = \sum_{k=t}^{\infty}\gamma^{k-t} \phi_{k+1}.
\end{equation}

\begin{figure}
    \centering
\begin{tikzpicture}[
    >=Latex, % Nicer arrow tips
    font=\small % Adjust font size if needed
  ]
  % --- Define constants ---
  \def\meanval{1.5} % Position of the mean cost (center point)
  \def\axismin{-1.5} % Left end of the axis
  \def\axismax{5.5} % Right end of the axis
  \def\distspread{0.8} % Controls the width of the distribution (like std dev)
  \def\distheight{1.8} % Controls the peak height of the distribution
  \def\labeloffset{2.5} % Adjusted Horizontal distance for main labels
  \def\labelvertoffset{1.8} % Adjusted Vertical position for main labels
  
  % VaR and CVaR positions (now on the RIGHT tail for cost)
  \def\varpos{2.15} % VaR - the boundary of the high-cost tail
  \def\cvarpos{3.3} % CVaR - mean of the shaded high-cost tail region
  
  % --- Draw Axis ---
  \draw [->, thick] (\axismin, 0) -- (\axismax, 0) node[above left, xshift=-1mm] {Cost};
  
  % Add ticks
  \foreach \x in {-1, 0, 1, ..., 5} {
      \draw (\x, 3pt) -- (\x, -3pt) node[below=4pt] {$\x$};
  }
  
  % --- Draw the distribution shape ---
  \draw[black, thick, domain=\meanval-3*\distspread:\meanval+3*\distspread, samples=100, smooth]
    plot (\x, { \distheight * exp( -((\x-\meanval)^2) / (2*\distspread^2) ) });
  
  % --- Shade the tail region 
  \fill[gray!20, domain=\varpos:\meanval+3*\distspread, samples=50, smooth]
    plot (\x, { \distheight * exp( -((\x-\meanval)^2) / (2*\distspread^2) ) }) -- (\meanval+3*\distspread, 0) -- (\varpos, 0) -- cycle;
  
  % --- VaR line (boundary) ---
  \draw [black!70!black, thick, dashed] (\varpos, -0.3) -- (\varpos, { \distheight * exp( -((\varpos-\meanval)^2) / (2*\distspread^2) ) + 0.2 }) 
    node[above, yshift=1mm] {$\text{VaR}_{\alpha}$};
  
  % --- CVaR line (expected value of tail) ---
  \draw [black!50!black, thick, dashdotted] (\cvarpos, -0.3) -- (\cvarpos, { \distheight * exp( -((\cvarpos-\meanval)^2) / (2*\distspread^2) ) + 0.2 }) 
    node[above, yshift=1mm] {$\text{CVaR}_{\alpha}$};
  
  % --- Mean line ---
  \draw [gray, dashed] (\meanval, -0.5) -- (\meanval, \distheight + 0.3) 
    node[above, yshift=1mm, black] {$Q(\bm{x}_t,\bm{u}_t) = \mathbb{E}[Z(\bm{x},\bm{u})]$};
  
\end{tikzpicture}
\caption{Distribution of costs for a given state-action pair, $Z(\bm{x}_t, \bm{u}_t)$, with Value-at-Risk (VaR$_{\alpha}$) denoting the cost threshold where the probability of exceeding it is $(1-\alpha)$, and Conditional Value-at-Risk (CVaR$_{\alpha}$) representing the expected cost conditional on exceeding VaR$_{\alpha}$ (i.e., the mean of the worst $(1-\alpha)$ fraction of outcomes).}
\label{fig:dist_rl}
\end{figure}
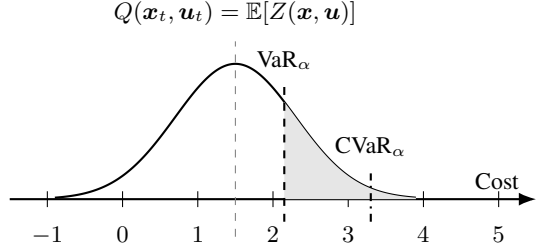

Notice that \eqref{eq:randomreturn} now treats the state-action value as a random variable, which must be represented as a distribution, compared to the Q-function.  
In other words, this random variable $Z^\pi(\bm{x}_t, \bm{u}_t)$ captures the full distribution over possible returns, with $Q^\pi(\bm{x}_t, \bm{u}_t) = \mathbb{E}[Z^\pi(\bm{x}_t, \bm{u}_t)]$ recovering the standard value function (see Figure~\ref{fig:dist_rl}). The distributional perspective generally motivates value-function approximation algorithms, using the application of the distributional Bellman operator,
\begin{equation*}
Z^\pi(\bm{x}_t, \bm{u}_t) \stackrel{d}{=} \phi(\bm{x}_t, \bm{u}_t, \bm{x}_{t+1}) + \gamma Z^\pi(\bm{x}_{t+1}, \pi(\bm{x}_{t+1})),
\end{equation*}
where $\bm{x}_{t+1} \sim p(\cdot|\bm{x}_t, \bm{u}_t)$ and $\stackrel{d}{=}$ denotes equality in distribution. This formulation captures the propagation of uncertainty through the RL problem, directly addressing the distributional nature of the original stochastic control problem presented in Section~\ref{sec:problem_setting}. 
Given that $Z^\pi$ defines a distribution rather than a function (as is true for $Q^\pi$), distributional RL methods often must account for learning a distribution. 

Early methods represented $Z^\pi$ as a probability distribution over a fixed, discrete support and learned their associated probabilities using cross-entropy to a target that is projected onto the support~\cite{rowland2018analysis}. These categorical distributional RL methods parameterize the return distribution using fixed support locations $z_1 < z_2 < \cdots < z_N$ with learnable probabilities $p_1, \ldots, p_N$. Since the distributional Bellman target generally has support misaligned with these fixed locations, a projection operator $\Pi_C$ redistributes the target's probability mass onto the nearest grid points via linear interpolation. The algorithm then minimizes the Kullback-Leibler divergence between this projected target and the current prediction.

Alternatively, quantile regression~\cite{dabney2018distributional} takes the converse approach: it fixes the probabilities (as uniform quantiles $\tau_i = i/N$) and learns adjustable support locations, avoiding the projection step entirely. Both categorical and quantile methods were later extended from tabular to function approximation settings using neural networks for TD learning~\cite{rowland2024analysis} and distributional Q-learning~\cite{yang2019fully}. More recently, \citet{bloor2025gaussian} use Gaussian processes to represent distributions over Q-value functions for a finite-horizon, semi-batch reactor control problem.

% Algorithms / PSE works
\subsection{Distributional RL in PSE}
The works by \citet{wu2023distributional} and \citet{kotechaMORL} provide examples in the PSE supply chain domain. They applied distributional RL concepts to optimize inventory management in multi-echelon supply chains. Recognizing that standard RL optimizing average profit (equivalent to minimizing average cost) could expose the system to significant risks (e.g., large stockouts or holding costs during demand spikes), they incorporated a CVaR constraint into their objective function. Using a hybrid derivative-free policy search algorithm, they evaluated policies based not only on expected profit but also on the Conditional Value at Risk (CVaR) of the profit distribution, estimated via MC rollouts. Their results demonstrated that the risk-sensitive policy learned via distributional RL significantly improves the worst-case performance (achieving a much higher CVaR, indicating better protection against large losses) compared to a policy optimizing only the expected profit, often with only a small trade-off in average performance. 
Similarly, \citet{bermudez2023distributional} propose a distributional RL algorithm for inventory management that further accounts for the distributional aspect of violating constraints (constraints are discussed further in Section~\ref{sec:constrained_rl}). 
These applications highlight the practical utility of distributional RL for developing more robust and reliable operational strategies in uncertain PSE contexts.

%%%%%%%%%%%%%%%%%%%%%%%%%%%%%%%%%
\section{Model-Based RL}\label{sec:model_based_rl}
The RL approaches discussed in Sections \ref{sec:model_free_methods}--\ref{sec:distributional_rl} operate directly on experience tuples $( \bm{x}_t, \bm{u}_t, \phi_t, \bm{x}_{t+1} )$ without explicitly modeling the system dynamics. While effective, these model-free methods often require extensive interactions with the environment to achieve satisfactory performance. For process systems with high operational costs, safety-critical constraints, or limited data availability, extensive exploration may be impractical or prohibitively expensive. 
We refer the reader to \cite{thebelt2022maximizing} for an overview of data acquisition challenges (and associated modeling challenges) in chemical engineering systems. 

Model-based RL offers an alternative paradigm that addresses these limitations by explicitly learning a model of the environment dynamics. By constructing a model $\hat{p}(\bm{x}_{t+1} | \bm{x}_t, \bm{u}_t)$ that approximates the true transition dynamics $p(\bm{x}_{t+1} | \bm{x}_t, \bm{u}_t)$, the agent can be given an `internal' model with which to plan, reminiscent of many control approaches. With an internal model, the agent can simulate potential future trajectories, evaluate the consequences of different control sequences, and reason about long-term outcomes without actually executing controls in the real environment. This planning capability enables more efficient learning and informed decision-making, significantly reducing the need for costly real-world interactions while maintaining or even improving performance.

\subsection{Model-Based RL Methodologies}
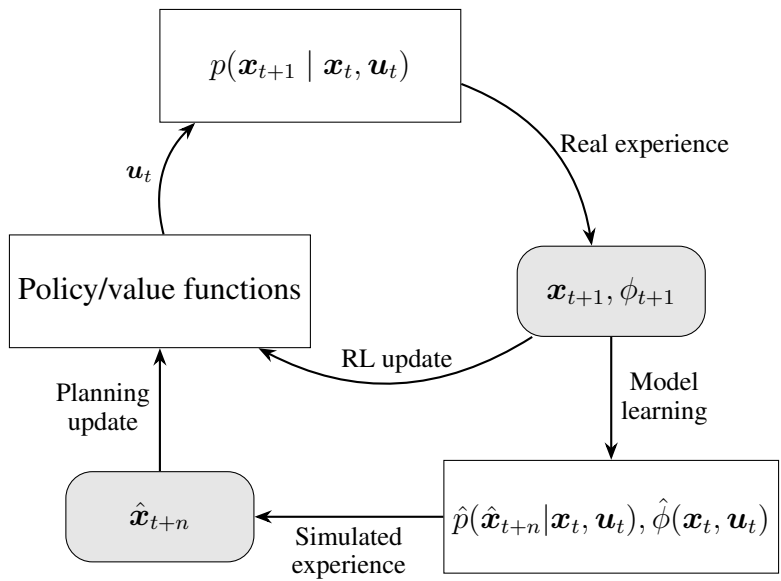
\begin{figure}
    \centering

\begin{tikzpicture}[
    box/.style={rectangle, draw, minimum width=4cm, minimum height=1.5cm, text centered, font=\large},
    roundbox/.style={rectangle, rounded corners=10pt, draw, fill=gray!20, minimum width=2.5cm, minimum height=1.2cm, text centered, font=\large},
    arrow/.style={-Stealth, thick}
]

% Define the main components
\node[box] (policy) at (-2,1) {Policy/value functions};
\node[roundbox] (realexp) at (4,1) {$\bm{x}_{t+1},\phi_{t+1}$};
\node[box] (env) at (0,4) {$p(\bm{x}_{t+1}\mid \bm{x}_t,\bm{u}_t)$};
\node[box] (model) at (4,-2) {$\hat{p}(\hat{\bm{x}}_{t+n} | \bm{x}_t, \bm{u}_t), \hat{\phi}(\bm{x}_t,\bm{u}_t)$};
\node[roundbox] (simexp) at (-2,-2) {$\hat{\bm{x}}_{t+n}$};

% Draw the rest of the connections
\draw[arrow] (env) to[bend left=30] node[right, align=center] {Real experience} (realexp);
\draw[arrow] (realexp) to[bend left=30] node[above, align=center] {RL update} (policy);
\draw[arrow] (simexp) to node[left, align=center] {Planning\\ update} (policy);
\draw[arrow] (realexp) to node[right, align=center] {Model\\learning} (model);
\draw[arrow] (model) to node[below, align=center] {Simulated \\experience} (simexp);
\draw[arrow] (policy) to[bend left=30] node[left, align=center] {$\bm{u}_t$} (env);
\end{tikzpicture}
\caption{Dyna Framework~\cite{dyna}. The policy and value functions are updated with both real experience (as in model-free RL) and with simulated experience using a dynamics $\hat{p}(\hat{\bm{x}}_{t+n} | \bm{x}_t, \bm{u}_t)$ and cost model $\hat{\phi}(\bm{x}_t,\bm{u}_t)$}
\label{fig:dyna}
\end{figure}

We highlight here the Dyna~\cite{dyna} architecture as a representative and foundational framework that leverages this planning capability by integrating model-based planning with model-free learning. Rather than treating these approaches as distinct alternatives, Dyna combines them by using a learned model to generate simulated experiences that supplement real interactions, accelerating the learning process without sacrificing the asymptotic performance of model-free methods. 
In other words, the RL agent interacts with the learned model (and optionally the real world) as its `environment,' while the learned model can be improved using the real-world interactions (Figure~\ref{fig:dyna}). 

The core Dyna algorithm alternates between real-world interaction to collect transition tuples, model learning to update the transition and cost models, direct RL to update the value function or policy based on real experience, and model-based planning to generate simulated experiences for further updates. Through these processes, Dyna exploits the model's planning capability to enhance learning.

The Dyna framework can be formalized using Q-learning as the model-free component (see \eqref{eq:qlearning_update}) for real experiences, and,
\begin{equation*}
Q^{\pi}(\bm{x}_t, \bm{u}_t) \leftarrow Q^{\pi}(\bm{x}_t, \bm{u}_t) + \alpha[\hat{\phi}_t(\bm{x}_t, \bm{u}_t) + \gamma \min_{\bm{u}_{t+1}} Q^{\pi}(\hat{\bm{x}}_{t+1}, \bm{u}_{t+1}) - Q^{\pi}(\bm{x}_t, \bm{u}_t)],
\end{equation*}
for simulated experiences, where $\hat{\bm{x}}_{t+1} \sim \hat{p}(\cdot | \bm{x}_t, \bm{u}_t)$.

While Dyna provides a foundational framework for integrating model-based and model-free learning, several alternative methodologies have been developed within the model-based reinforcement learning paradigm. PILCO~\cite{deisenroth2011pilco} utilizes Gaussian processes to construct probabilistic dynamics models that enable uncertainty-aware policy optimization through probabilistic inference. Model-Based Policy Optimization ~\cite{janner2019trust} employs truncated model rollouts in conjunction with model-free policy optimization to address model exploitation and distributional shift. World Models~\cite{ha2018world} construct compact latent representations of environmental dynamics, facilitating policy training entirely within the learned model's simulated environment. These methodologies share the fundamental principle of exploiting learned dynamics models to enhance sample efficiency, yet differ substantially in their approaches to model representation, uncertainty quantification, and the integration of model-based planning with model-free optimization.
% MB approaches in PSE
\subsection{Model-based RL in PSE}
Beyond using learned models purely for generating simulated data as in Dyna, many model-based RL techniques directly incorporate the known or learned model structure into the value function or policy updates. This is particularly common in approaches derived from Approximate Dynamic Programming (ADP). For example, \citet{model_based_kim2020} developed a model-based method using Globalized Dual Heuristic Programming  for finite-horizon optimal control of nonlinear control-affine systems. Their approach leverages the known system dynamics and cost function structure to iteratively learn approximations of the value, costate (value gradient), and policy functions derived from the underlying Hamilton-Jacobi-Bellman equation. By employing Deep Neural Networks as function approximators and incorporating stabilization techniques from deep RL such as replay buffers and target networks, they demonstrated robust performance and improved scalability compared to methods using shallower networks, particularly for higher-dimensional systems with uncertainties. This work exemplifies how explicit model knowledge can be integrated with deep learning within an RL framework to solve structured optimal control problems encountered in PSE.

Among the various advanced process control strategies employed in the PSE community, MPC is the most established and has become central to chemical process industry, power and energy systems, as well in robotics and aerospace applications. Model-based RL shares the fundamental principle of using a model to plan over future horizons, but differs in a key way. MPC repeatedly solves mathematical programs defined on a given time horizon structure, exploiting available descriptions of process dynamics and constraints; while model-based RL learns both dynamics models and policy function approximations from data to enable cheap online execution through prediction. Data-driven MPC \cite{BRADFORD2020106844,PILARIO202519} approaches represent a natural bridge between these paradigms, incorporating learned models while maintaining the framework familiar to control engineers. MPC has a major advantage in that the Karush-Kuhn-Tucker (KKT) optimality conditions provide means to explicitly account for and handle constraints imposed on operations. This is a major challenge posed to MDP solvers and is the primary focus of the constrained RL subfield. 

%%%%%%%%%%%%%%%%%%%%%%%%%%%%%%%%%
\section{Constrained RL}\label{sec:constrained_rl}

Process systems engineering applications frequently necessitate strict enforcement of operational constraints to ensure safety (e.g., temperature limits), product quality (e.g., concentration bounds), environmental regulations (e.g., emission caps), and/or physical feasibility (e.g., inventory capacities, valve limits). Standard RL algorithms, which typically optimize an unconstrained cumulative cost or cost objective (Eq. \ref{eq:finite_problem} or its infinite horizon equivalent), do not explicitly guarantee that these constraints will be satisfied during either learning or deployment phases. Applying unconstrained RL policies directly to physical systems can therefore lead to unsafe operation (e.g., during exploration), off-spec products, or infeasible controls, representing a significant barrier to the adoption of RL in constraint-critical industrial settings.

Safety requirements in PSE can be characterized along two key dimensions. The type of safety distinguishes between statistical safety (constraints satisfied in expectation) and hard safety (constraints satisfied at every time step). The phase of concern distinguishes between training safety (during exploration) and deployment safety (for converged policies). PSE applications typically require hard safety during both phases, presenting challenges for standard constrained RL approaches that primarily address statistical safety for converged policies.

\subsection{Constrained Markov Decision Processes}
The standard theoretical framework for this problem is the Constrained Markov Decision Process (CMDP)~\citep{cmdp_book}. A CMDP extends the standard MDP tuple $\mathcal{M} = \langle \mathcal{X}, \mathcal{U}, p, \phi, \gamma\rangle$ to include $m$ auxiliary cost functions $c_i: \mathcal{X} \times \mathcal{U} \times \mathcal{X} \rightarrow \mathbb{R}$ and corresponding thresholds $\kappa_i \in \mathbb{R}$. The objective is to find a policy $\pi$ that minimizes the primary expected cumulative cost $J(\pi)$ subject to constraints on the expected cumulative auxiliary costs $J_{C_i}(\pi)$:

\begin{align}
\min_{\pi} \hspace{2mm} &J(\pi) = \mathbb{E}_{p_\pi}\left[\sum_{t=0}^{\infty} \gamma^t \phi(\bm{x}_t, \bm{u}_t, \bm{x}_{t+1})\right] \nonumber\\
\text{s.t.}\hspace{0.5cm}
&J_{C_i}(\pi) = \mathbb{E}_{p_\pi}\left[\sum_{t=0}^{\infty} \gamma^t c_i(\bm{x}_t, \bm{u}_t, \bm{x}_{t+1})\right] \leq \kappa_i, \quad \forall i \in \{1,\ldots,m\} \nonumber\\
& p_\pi(\bm{\tau}) = p_0(\bm{x}_0) \prod_{t=0}^{\infty} \pi(\bm{u}_t|\bm{x}_t) p(\bm{x}_{t+1}|\bm{x}_t,\bm{u}_t) \nonumber.
\end{align}

\begin{figure}[htbp]
\centering
\includesvg[width=0.4\textwidth]{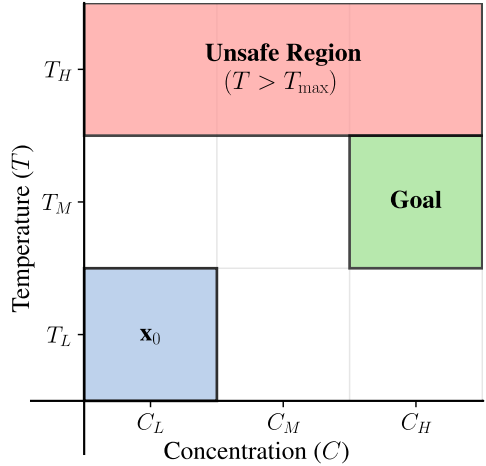}
\caption{Process control grid world with a maximum temperature constraint. The agent must reach the target region while avoiding the unsafe high-temperature zone.}
\label{fig:process_control_constrained}
\end{figure}

Figure~\ref{fig:process_control_constrained} provides an illustration of a CMDP in a process control context. The crucial constraint is to avoid the `Unsafe Region', where the temperature exceeds a predefined maximum, $T_{max}$. This safety requirement is captured by an auxiliary cost function, $c^1(\bm{x}_t, \bm{u}_t, \bm{x}_{t+1})$, which assigns a high penalty if the subsequent state $\bm{x}_{t+1}$ falls within this unsafe high-temperature zone (i.e., $T_H$ region), and a low or zero cost otherwise. The CMDP objective is then to find a policy $\pi$ that minimizes the expected cumulative primary cost $J(\pi)$ while ensuring that the expected cumulative constraint cost $J_C^1(\pi)$ remains below a specified threshold $\kappa^1$. Importantly, this CMDP formulation provides only statistical safety guarantees. The constraint handling does not prohibit individual constraint violations, but simply ensures the expected constraint cost remains below a certain value over many episodes.

\subsection{Solution Approaches}
Common approaches for solving CMDPs adapt standard RL algorithms with tools from constrained optimization, such as Lagrangian relaxation methods that convert the constrained problem into an unconstrained problem using Lagrange multipliers $\lambda_i \geq 0$. The optimization alternates between minimizing the Lagrangian,
\begin{equation*}
    \mathcal{L}(\pi, \bm{\lambda}) = J(\pi) + \sum_{i=1}^m \lambda_i (J_{C_i}(\pi) - \kappa_i),
\end{equation*}
with respect to the policy $\pi$, and maximizing it with respect to the multipliers $\bm{\lambda}$ (typically via gradient ascent). 
The first step is analogous to standard RL on a modified cost $\phi + \sum \lambda_i c_i$ and can be addressed by applying the algorithms described in Sections~\ref{sec:model_free_methods}-\ref{sec:model_based_rl} to minimize $\mathcal{L}(\pi, \bm{\lambda})$ instead of $J(\pi)$. 
Algorithms such as Constrained Policy Optimization (CPO) \citep{cpo} build upon this principle, often integrating it with policy gradient or actor-critic methods.

Penalty methods incorporate constraints directly into the cost function as penalties for violations, e.g., 
\begin{equation*}
    \tilde{\phi}(\bm{x}_t,\bm{u}_t,\bm{x}_{t+1}) = \phi(\bm{x}_t,\bm{u}_t,\bm{x}_{t+1})+ \sum \beta_i  \max(0, c_i(\bm{x}_t,\bm{u}_t,\bm{x}_{t+1})- \kappa_i),
\end{equation*}
or similar pathwise penalties. This transforms the problem into an unconstrained one solvable by standard RL, but finding appropriate penalty coefficients $\beta_i$ can be difficult, and strict constraint satisfaction is not guaranteed. Notable algorithms implementing these approaches include Reward Constrained Policy Optimization (RCPO)~\cite{tessler2018reward}, which uses a Lagrangian formulation to balance reward maximization with constraint satisfaction, and PPO-Lagrangian~\cite{ray2019benchmarking}, which extends Proximal Policy Optimization with dual gradient ascent on Lagrange multipliers, 

\subsection{Constrained RL in PSE}
This section identifies a  limitation in standard CMDP formulations for PSE safety requirements. PSE applications necessitate explicit pathwise or terminal constraint satisfaction rather than expected performance guarantees. The cited works demonstrate an evolution toward specialized approaches that embed predictive models, uncertainty quantification, and constrained optimization directly within RL frameworks to provide the real-time safety guarantees that PSE applications demand.

\citet{pan2021constrained} introduced an oracle-assisted Q-learning method where learned functions predict maximum future constraint violations. These functions are then utilised to enforce chance constraint satisfaction. Control selection involves maximizing the Q-value subject to these predictions plus tunable backoffs, providing foresight within a value-based framework. \citet{petsagkourakis2022chance} proposed a Chance Constrained Policy Optimization (CCPO) algorithm that guarantees satisfaction of joint chance constraints with high probability, addressing the critical limitation that prior methods only ensured constraint satisfaction in expectation, by computing constraint backoffs simultaneously with the feedback policy using Bayesian optimization over the empirical cumulative distribution function. Similarly, \citet{mowbray2022safe} used Gaussian Processes (GPs) to model dynamics and explicitly quantify uncertainty. They derived deterministic surrogate constraints using backoffs based on the GP's predictive variance (tuned via Bayesian Optimization) and modified the cost function for policy gradient optimization to enforce chance constraints and penalize high model uncertainty. Alternatively, \citet{BURTEA2024108518} tackled known pathwise constraints in continuous-control Q-learning by performing control selection via direct constrained optimization over the learned Q-function network at each step, using tools such as OMLT~\cite{ceccon2022omlt} to guarantee feasible controls during deployment.

These examples illustrate a trend towards embedding predictive constraint models, probabilistic uncertainty quantification, or explicit constrained optimization solvers within the RL framework. Additionally, control-theoretic concepts such as control barrier functions \cite{marvi2021safe}, control invariant sets, and neural Lyapunov functions \cite{choi2020reinforcement} offer pathways to pathwise safety guarantees and formal stability certificates for the learned policies. This integration aims to provide the stronger safety and feasibility guarantees needed for practical deployment of RL in constraint-critical PSE applications, moving beyond restrictions solely on expected cumulative costs.

%%%%%%%%%%%%%%%%%%%%%%%%%%%%%%%%%
\section{Offline RL}\label{sec:offline_rl}
The RL methods discussed to this point operate under the assumption that the agent can freely interact with the environment during training, collecting new experiences through exploration. In reality, for many PSE applications, active exploration may be prohibitively expensive, potentially dangerous, or simply impractical due to operational constraints~\cite{thebelt2022maximizing}. Offline RL (also known as batch RL) addresses this limitation by learning optimal policies exclusively from previously collected datasets, without any environment interaction during the training phase.

\subsection{Problem Formulation}
In the offline RL setting, we are provided with a fixed dataset $\mathcal{D} = \{( \bm{x}^i_t, \bm{u}^i_t, \phi^i_t, \bm{x}^i_{t+1} )\}_{i=1}^N$ of transitions collected under some behavior policy (or set of policies) $\pi_\beta$. This dataset might represent historical operating data from a chemical plant, batch process records, or simulation results under various control strategies. The overall goal is to learn an optimal policy $\pi^*$ that maximizes expected return when deployed, without the ability to collect additional data through exploration. Formally, the offline RL problem is identical to~\eqref{prob:inf_horizon} except the objective function is subject to the constraint that learning must occur using only the fixed dataset $\mathcal{D}$.

\subsection{Distributional Shift and Extrapolation Error}
The fundamental challenge in offline RL stems from the distributional shift between the behavior policy that generated the dataset and the target policy being learned. Traditional off-policy algorithms such as Q-learning and DDPG assume the ability to collect new data to correct estimation errors. Without this capability for active learning, these algorithms often suffer from extrapolation errors, e.g., where the learned Q-function approximator must estimate values for state-action pairs not present in the dataset.

The extrapolation error often manifests as systematic underestimation of Q-values for out-of-distribution controls, leading to exploitative policies that perform poorly when deployed~\cite{kumar2020conservativeqlearningofflinereinforcement}. This occurs because standard Bellman updates (Equation~\eqref{eq:qlearning_update}) involve minimization over controls.
As training progresses, the policy increasingly selects controls with optimistically overestimated Q-values, further exacerbating the problem.

\subsection{Conservative Q-Learning}
\begin{figure}
    \centering
    \includegraphics[width =0.4\textwidth]{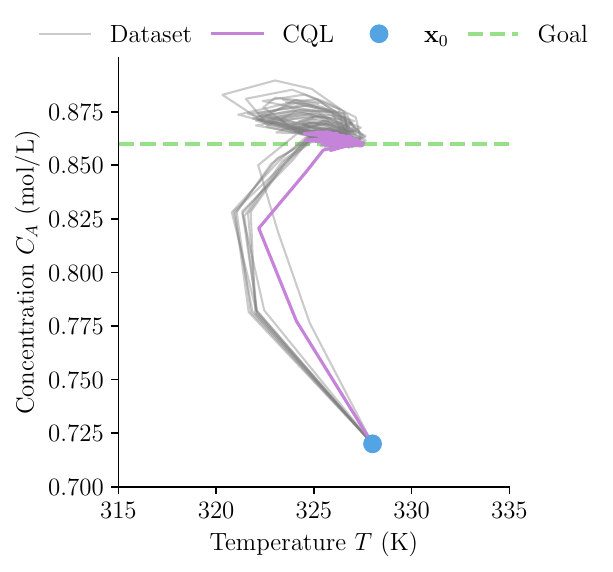}
    \caption{Conservative Q-learning with the dataset shown in grey generated from a policy trained by DQN. The resulting trajectory of CQL shown in pink.}
    \label{fig:CQL}
\end{figure}

Conservative Q-Learning~\cite{kumar2020conservativeqlearningofflinereinforcement} (CQL) addresses the above extrapolation error by explicitly overestimating costs for out-of-distribution controls while underestimating costs for controls observed in the dataset (see Figure \ref{fig:CQL}). This ensures the agent remains conservative by avoiding actions with uncertain consequences. CQL modifies the standard Bellman error minimization objective with a conservative regularization term,
\begin{equation*}
\begin{aligned}
L_{CQL}(\omega) = \alpha \mathbb{E}_{\bm{x}_t \sim \mathcal{D}} \Bigg[&\mathbb{E}_{\bm{u}_t \sim \mathcal{D}(\cdot|\bm{x}_t)}[Q^{\pi}_{\omega}(\bm{x}_t, \bm{u}_t)] 
- \log \sum_{\bm{u}_t} \exp(-Q^{\pi}_{\omega}(\bm{x}_t, \bm{u}_t))\Bigg] \\
&+ \mathbb{E}_{( \bm{x}_t, \bm{u}_t, \phi, \bm{x}_{t+1} ) \sim \mathcal{D}} \left[ \left(Q^{\pi}_{\omega}(\bm{x}_t, \bm{u}_t) - \left(\phi + \gamma \min_{\bm{u}_{t+1}} Q^{\pi}_{\omega'}(\bm{x}_{t+1}, \bm{u}_{t+1})\right) \right)^2 \right],
\end{aligned}
\end{equation*}
where $\omega'$ represents the parameters of a target network, and $\alpha$ is a coefficient controlling the level of conservatism. The first term underestimates costs for dataset controls while overestimating costs for all actions, particularly out-of-distribution ones (soft-minimum via a log sum exponent). When minimizing this objective, the negative sign on the soft-minimum causes Q-values for out-of-distribution controls to increase (higher cost estimates), while dataset Q-values decrease (lower cost estimates). This formulation effectively provides a conservative cost estimate that discourages exploration into uncertain regions of the control space.

\subsection{Off-policy Evaluation}
After learning policies from offline datasets, PSE settings frequently require estimation of a policy's performance before deployment given only historical data. This off-policy evaluation (OPE) problem is particularly relevant in industrial settings where deploying an untested policy is inherently risky and expensive.

The OPE problem seeks to estimate the expected return $J(\pi) = \mathbb{E}_{\pi}[\sum_{t=0}^{T-1} \gamma^t \phi_t]$ for a target policy $\pi$ using only data collected under behavior policy $\pi_\beta$. Importance sampling provides the foundational approach~\cite{precup2000eligibility}, weighting observed returns by the likelihood ratio between policies. This re-weighting of returns by the policy likelihood ratio effectively corrects for the discrepancy between the data-generating policy and the target policy, thus forming an estimate for the OPE problem,
\begin{equation}
\hat{J}_{IS}(\pi) = \frac{1}{M}\sum_{m=1}^{M} G_m \prod_{t=0}^{T-1} \frac{\pi(\bm{u}^m_t|\bm{x}^m_t)}{\pi_\beta(\bm{u}^m_t|\bm{x}^m_t)}
\end{equation}
where $G_m$ is the return from episode $m$. 

However, fundamental challenges limit OPE's practical applicability in PSE. Most industrial controllers are deterministic. This means $\pi_\beta$ places all probability mass on a single control. If the learned policy $\pi$ proposes a different control, the importance weight becomes zero, yielding uninformative estimates whenever policies disagree. Additionally, OPE assumes stationary dynamics; if the process has shifted due to equipment degradation or operational changes, historical data no longer reflects the deployment environment. While methods like weighted importance sampling~\cite{mahmood2014weighted} and doubly robust estimators~\cite{dudik2011doublyrobustpolicyevaluation} reduce variance, they do not resolve these problems. These limitations suggest OPE requires further methodological progress by the PSE community before it can be used robustly.
 
\subsection{Offline RL in PSE}
Further addressing the practical implementation challenges of RL in process control, alternative strategies leveraging offline data or incorporating significant domain knowledge have been explored. \citet{mcclement2022meta} proposed a meta-RL approach where an agent is trained entirely offline across a distribution of simulated process dynamics (specifically FOPTD systems) to rapidly tune PI controllers online. This method utilizes a recurrent neural network structure to learn system context from online process data in a model-free manner after the offline training phase, thereby improving sample efficiency for adaptation to new systems within the learned distribution. While trained offline (potentially using privileged model information for the critic component during training) the deployed agent adapts online using only standard process measurements, though data augmentation may be required for systems significantly outside the training distribution. In contrast, \citet{patel2023practical} focused on enhancing the safety, speed, and explainability of online RL implementation through a systematic problem formulation methodology applied before utilizing standard RL algorithms such as DDPG. This approach incorporates domain knowledge explicitly through techniques such as state-space reduction, control adjustments (suppression, restriction, enforcement based on proximity to constraints), and curated cost functions, thereby reducing the burden on the RL algorithm itself and minimizing risky exploration. While primarily aimed at structuring online learning, this method also acknowledges the potential for offline retraining using the collected, structured online interaction data if initial online performance is insufficient. \citet{gil2025reinforcement} demonstrated a behavior cloning approach for pH control in open photobioreactors, where a DDPG agent was trained offline using trajectories from a nominal PID controller and subsequently fine-tuned online daily, achieving a pioneering real-world deployment of RL-based control for such highly nonlinear, disturbance-prone bioprocesses with reduced control effort and improved tracking performance. Similarly, \citet{wang2025offline} applied Implicit Q-Learning (IQL) to train control policies from historical bioprocess data for fed-batch optimization, followed by online fine-tuning using TD3 with mixed experience replay, demonstrating superior performance compared to behavioral cloning while addressing the data scarcity and safety risks inherent in purely online training approaches. Offline RL has also been applied to flatness control in steel strip rolling \cite{deng2023offline}, where the authors used an ensemble of Q-functions within a TD3-BC framework to learn from factory data, addressing uncertainty and distributional shift without requiring a simulator. \citet{Park2026offline} applied DQN-based Cal-QL with LSTM networks for industrial dividing wall column temperature control, demonstrating successful offline pre-training on historical operational data followed by online fine-tuning for real-world deployment, addressing the discrete action distributions and time-series structure typical of manual operations in complex multi-component distillation systems.

A complementary but underexplored area in PSE is OPE, which addresses the critical need to assess policy performance using only historical data before deployment. Given the availability of historical plant data and controllers in industrial processes and the high costs associated with deploying untested control policies, OPE methods could provide substantial value by enabling pre-deployment assessment of RL-based controllers using existing operational data, thereby reducing deployment risks while leveraging the exisiting wealth of historical process information.

%%%%%%%%%%%%%%%%%%%%%%%%%%%%%%%%%%%%%%%%%%%%%%%%
\section{Goal-Conditioned Reinforcement Learning}
\label{sec:gcrl}
While standard RL formulations optimize policies for a single, fixed objective defined by the (cumulative) reward function, many PSE applications require agents that can adapt to multiple objectives or operate across varying setpoints (e.g. multi-level problems) and operational targets. 
For instance, consider process control, where it is desirable to learn a setpoint-dependent policy that drives the system to a given setpoint. 
Goal-Conditioned RL (GCRL) addresses this limitation by generalising the RL framework to include some goal to be achieved, making it particularly relevant for PSE applications where flexibility and adaptability are required~\cite{liu2022goalconditionedreinforcementlearningproblems}.

\subsection{Problem Formulation}

GCRL extends the standard MDP framework by augmenting it with an explicit goal space. A Goal-Augmented MDP (GA-MDP) is defined as the tuple $\mathcal{M}_G = \langle \mathcal{X}, \mathcal{U}, \mathcal{G}, p, \phi_g, \gamma, p_g, \varphi \rangle$, where $\mathcal{G}$ represents the goal space describing desired task outcomes, $p_g$ denotes the distribution over goals, $\varphi: \mathcal{X} \rightarrow \mathcal{G}$ is a mapping function from states to goals, and $\phi_g: \mathcal{X} \times \mathcal{U} \times \mathcal{G} \rightarrow \mathbb{R}$ is the goal-conditioned cost function~\cite{liu2022goalconditionedreinforcementlearningproblems}.

\begin{figure}
    \centering
    \includesvg[width=0.8\linewidth]{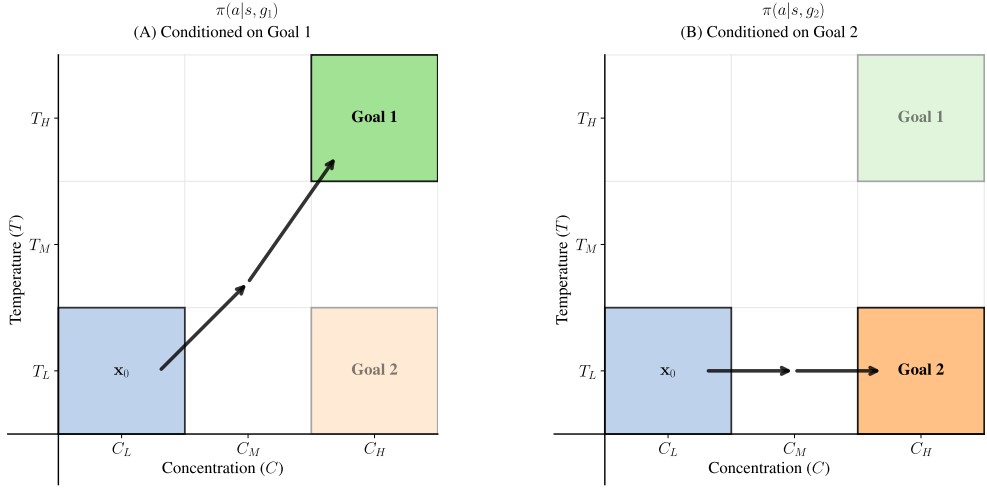}
    \caption{Illustration of a goal-conditioned policy in the process control gridworld.}
    \label{fig:gcrl}
\end{figure}

In this framework, the agent receives both the current state $\bm{x}_t$ and a desired goal $\bm{g}$ as its input at time $t$, and it must learn a goal-conditioned policy $\pi(\bm{u}_t|\bm{x}_t, \bm{g})$, as shown in Figure \ref{fig:gcrl}. The objective for GCRL therefore becomes,
\begin{align}
\min_{\pi} \hspace{2mm}&\mathbb{E}_{\bm{g} \sim p_g, \bm{\tau} \sim p_{\pi}} \left[\sum_{t=0}^{T-1} \gamma^t \phi_g(\bm{x}_t, \bm{u}_t, \bm{x}_{t+1}, \bm{g})\right],
\end{align}
where the expectation is taken over both the goal distribution and the trajectory distribution induced by the policy.

The goal-conditioned framework introduces three fundamental concepts that characterize the interaction between agent and environment: (1) a  \textit{desired goal} represents the target outcome specified by the task or environment, which may be provided externally or generated through curriculum learning strategies; (2) an \textit{achieved goal} corresponds to the goal(s) that the agent has actually reached in its current state, typically defined as $\varphi(\bm{x}_t)$; and (3) a \textit{behavioral goal} denotes the specific objective being pursued during a particular episode, which may differ from the original desired goal due to goal substitution or hindsight relabeling techniques~\cite{liu2022goalconditionedreinforcementlearningproblems}.
%%%%%%%%%%%%%%%%%%%%%%%%%
%%%%%%%%%%%%%%%%%%%%%%%%%

\subsection{Sparse Reward Structure and Exploration Challenges}

GCRL environments typically employ sparse, binary reward structures to reflect the goal-oriented nature of tasks. The reward function is commonly defined as,
\begin{equation}\label{eq:gcrl}
\phi_g(\bm{x}_t, \bm{u}_t, \bm{x}_{t+1},\bm{g}) = \begin{cases}
0 & \text{if } \|\varphi(\bm{x}_{t+1}) - \bm{g}\| \leq \epsilon \\
+1 & \text{otherwise}.
\end{cases}
\end{equation}
In other words, the actor receives 0 cost if the goal $\bm{g}$ is achieved and -1 if it is not, where goal achievement is determined by whether the achieved goal lies within tolerance $\epsilon$ of the desired goal~\cite{plappert2018multigoalreinforcementlearningchallenging}. This sparse reward structure creates significant exploration challenges, as agents can receive minimal learning signal during initial training phases when successfully achieving goals is rare. 
Intuitively, if the agent does not achieve any goals, $\phi_g=-1$ for all $\bm{x}_t$ in \eqref{eq:gcrl}, making it difficult to compute meaningful gradients, or otherwise improve the current policy. 

The sparsity problem is particularly acute in process systems applications where precise setpoint tracking or specific operational targets must be achieved. 
A reasonable solution is to `densify' the reward function, known as dense reward shaping, such as by incorporating some distance-based functions rather than the indicator function in \eqref{eq:gcrl}. 
This reward shaping can alleviate the sparsity-induced learning challenge, but may introduce artificial local optima or bias the learned policy toward suboptimal strategies~\cite{plappert2018multigoalreinforcementlearningchallenging}. Therefore, specialized techniques for handling sparse rewards in goal-conditioned settings become essential for practical PSE applications.

\subsection{Universal Value Function Approximators}

Universal Value Function Approximators (UVFAs) extend standard value functions to be conditioned on both states and goals, enabling a single function approximator to generalize across the entire goal space~\cite{pmlr-v37-schaul15}. More precisely, instead of learning separate value functions for each possible goal, UVFAs learn a unified function $V_{\theta}(\bm{x}, \bm{g})$ or $Q_{\theta}(\bm{x}, \bm{u}, \bm{g})$ parameterized by $\theta$.

The UVFA approach employs a factorized architecture, where states and goals are mapped to embedding vectors through learned functions $\varphi_x(\bm{x}, \bm{u})$ and $\varphi_g(\bm{g})$, respectively. These embeddings are subsequently used as inputs for a learned function $h$ to produce value estimates:
\begin{equation*}
Q_{\omega}(\bm{x}, \bm{u}, \bm{g}) = h(\varphi_x(\bm{x}, \bm{u}), \varphi_g(\bm{g}))
\end{equation*}

This factorization enables sharing of learned representations between similar states and goals, improving sample efficiency and generalization capabilities~\cite{pmlr-v37-schaul15}. The embedding functions can exploit structure in both state and goal spaces, with symmetric architectures being particularly effective when goals are defined in the same space as states, as is common in setpoint tracking applications.

\subsection{Hindsight Experience Replay}

Hindsight Experience Replay (HER) addresses the sparse reward problem by reinterpreting unsuccessful experiences as successful demonstrations for alternative goals~\cite{plappert2018multigoalreinforcementlearningchallenging}. The fundamental insight underlying HER is that any trajectory represents a successful demonstration of reaching the goals that were actually achieved during execution, regardless of the original intended goal.

Consider an episode executed with a single desired goal $\bm{g}$ that results in trajectory $\tau = (\bm{x}_0, \bm{u}_0, \ldots, \bm{x}_T)$. For each transition $(\bm{x}_t, \bm{u}_t, \bm{x}_{t+1})$ in this trajectory, HER generates additional training experiences by relabeling with alternative goals $\bm{g}'$ sampled from the set of achieved goals later in the episode,
\begin{equation*}
(\bm{x}_t, \bm{u}_t, \bm{g}', \phi_{\bm{g}'}(\bm{x}_t, \bm{u}_t, \bm{x}_{t+1}, \bm{g}'), \bm{x}_{t+1}).
\end{equation*}

The reward is recomputed according to a new goal $\bm{g}'$, typically resulting in positive feedback in terms of the reward function for transitions that would otherwise be treated as not being successful in achieving their original goal $\bm{g}$. This relabeling strategy dramatically increases the density of positive learning signals, while preserving the semantic meaning of the original task structure.

\subsection{Goal-Conditioned RL in PSE}

Goal-conditioned RL has proven particularly effective in many PSE applications due to its ability to adapt to multiple operational objectives and varying setpoints, addressing the complex, multi-objective nature of industrial processes. \citet{espinel2025reinforcement} demonstrated GCRL effectiveness in bioprocess control, specifically for multi-setpoint tracking in microbial consortia using multiplicative reciprocal saturation functions. Their approach overcomes limitations of traditional quadratic cost functions by ensuring any single reference deviation significantly reduces overall reward, guiding agents toward simultaneous objective satisfaction. The bioprocess application involved cybergenetic growth control through optogenetic mechanisms, with domain randomization enabling robust policies under varying conditions. \citet{lawrence2025viewlearningrobustgoalconditioned} integrated GCRL with model predictive control, using goal-conditioned value functions as terminal costs within robust MPC formulations~. This hybrid approach provides both RL adaptability and MPC safety guarantees for industrial deployment. Finally, \citet{li2023multi} developed goal-oriented multi-objective RL for fermentation processes using adaptive thresholded lexicographic ordering to balance conflicting objectives, demonstrating improved Pareto front approximation compared to traditional weighted approaches.

The practical advantages of the GRCL framework include robustness under process uncertainties through domain randomization, transfer learning capabilities enabling rapid adaptation to new setpoints without complete retraining, and integration with constraint handling mechanisms ensuring safety across different operational modes. These capabilities make GCRL particularly valuable for multi-product facilities where equipment must operate under varying specifications depending on manufacturing requirements.

%%%%%%%%%%%%%

\section{Synthesis and Outlook}
\label{sec:conc_future}
In this review, we survey the application of RL methods to PSE problems, revealing a strong focus on actor-critic algorithms for continuous control problems. As summarized in Table~\ref{tab:rl_overview}, approaches like DDPG, PPO, and TD3 are frequently applied to a range of challenges, from the setpoint scheduling of chemical reactors to the regulatory control of tank systems.
Despite successes in simulation, a persistent gap exists between these studies and practical industrial deployment, perhaps owing to the prohibitive sample complexity of many RL algorithms. However, this challenge also presents an opportunity: rather than treating RL algorithms as generic, model-agnostic tools, their data efficiency can be significantly improved by exploiting the inherent structure of PSE problems, or by leveraging offline RL techniques to warm-start the policy. 

This integration of domain knowledge is exemplified by the specialized RL subfields discussed herein. For instance, partial process models can be incorporated via Model-Based RL (Section~\ref{sec:model_based_rl}), a paradigm that shares conceptual similarities with Model Predictive Control~\cite{rawlings2020model} in its use of a dynamics model for planning, to reduce costly real-world interactions. Ensuring safety, however, remains paramount. While known operational limits can be encoded using the frameworks of Constrained RL (Section~\ref{sec:constrained_rl}) , significant research is still required in this area to provide the explicit guarantees (i.e., satisfying hard constraints) necessary for deployment in safety-critical systems. Distributional RL (Section~\ref{sec:distributional_rl}) enables risk-aware decision-making that accounts for the distributional nature of returns rather than merely their expected values, allowing operators to explicitly account for outcome variability and tail risks when making control decisions. Furthermore, historical plant data, a significant structural asset, can be leveraged through Offline RL (Section~\ref{sec:offline_rl}) to mitigate the risks and expense of online exploration. Even the objectives for RL can be structured for flexibility, as shown by Goal-Conditioned RL (Section~\ref{sec:gcrl}), which enables a single policy to adapt to varying operational targets. It should be noted that many of the algorithms mentioned throughout the review are available in software packages made by the RL community, many of which are advanced and robust.

This perspective both surveys and clarifies the potential role of RL within the established process control hierarchy. For well-characterized systems, RL is well-suited to the setpoint scheduling layer. Here, the problem is structured such that proven regulatory controllers handle low-level stability, while an RL agent optimizes high-level economic objectives against stochastic market forces~\cite{bloor2025hierarchicalrlmpcdemandresponse}. In contrast, for poorly understood systems where first-principles models are deficient, the model-free methods detailed in Section~\ref{sec:model_free_methods} provide a powerful framework for learning control policies directly from process data. 
Algorithms from RL may also play a role in data-driven approaches that integrate hierarchical decision making, or that bridge decision making across multiple levels of the hierarchy~\cite{tsay2019110th}.

Our outlook of RL intuitively aligns with the principles of systems thinking that are central to PSE. The successful application of RL in solving PSE challenges will therefore likely stem not from directly replacing classical scheduling and control, but from augmenting it with adaptive, data-driven policies that are intelligently integrated and rigorously tailored to the structured and constrained nature of the overall process system.

\section*{Acknowledgments}
MB would like to acknowledge funding provided by the Engineering \& Physical Sciences Research Council, United Kingdom through grant code  EP/W524323/1. CT gratefully acknowledges funding from a BASF/Royal Academy of Engineering Senior Research Fellowship.

%Bibliography
\bibliographystyle{unsrtnat}
\bibliography{bibliography}

\end{document}